\documentclass[%
 aip,
 pof,%
 amsmath,amssymb,
 preprint,%
]{revtex4-1}

\usepackage{caption}
\usepackage{graphicx}% Include figure files
\usepackage{dcolumn}% Align table columns on decimal point
\usepackage{bm}% bold math
\usepackage{subfig}
\usepackage{setspace}
\usepackage{tabularx,ragged2e,booktabs,caption}
\newcommand{\ra}[1]{\renewcommand{\arraystretch}{#1}}

\providecommand\bcdot{\boldsymbol{\cdot}}

\newcommand\p{\ensuremath{\partial}}
\DeclareTextFontCommand\textsfi{\usefont{OT1}{cmss}{m}{sl}}
\DeclareMathAlphabet\mathsfi            {OT1}{cmss}{m}{sl}
\DeclareTextFontCommand\textsfb{\usefont{OT1}{cmss}{bx}{n}}
\DeclareMathAlphabet\mathsfb            {OT1}{cmss}{bx}{n}
\DeclareTextFontCommand\textsfbi{\usefont{OT1}{cmss}{m}{sl}}
\DeclareMathAlphabet\mathsfbi            {OT1}{cmss}{m}{sl}

\begin{document}

\preprint{Phys. Fluids}

\title[BRIEF of Stokes flow]{Boundary regularized integral equation formulation (BRIEF) of Stokes flow}

\author{Q. Sun}
 \affiliation{Department of Mechanical Engineering, National University of Singapore, 10 Kent Ridge Crescent, 119260, Singapore}

\author{E. Klaseboer}%
 \affiliation{Institute of High Performance Computing, 1 Fusionopolis Way, 138632, Singapore}

\author{B. C. Khoo}
 \affiliation{Department of Mechanical Engineering, National University of Singapore, 10 Kent Ridge Crescent, 119260, Singapore}

\author{D. Y. C. Chan}
 \email{D.Chan@unimelb.edu.au.}
 \affiliation{Department of Mechanical Engineering, National University of Singapore, 10 Kent Ridge Crescent, 119260, Singapore}
 \affiliation{Institute of High Performance Computing, 1 Fusionopolis Way, 138632, Singapore}
 \affiliation{Department of Mathematics and Statistics, The University of Melbourne, Parkville 3010 VIC Australia}
 \affiliation{Department of Chemistry and Biotechnology, Swinburne University of Technology, Hawthorn 3122 VIC Australia}

%\date{\today}% It is always \today, today,
             %  but any date may be explicitly specified

\begin{abstract}
Single-phase Stokes flow problems with prescribed boundary conditions can be formulated in terms of a boundary regularized integral equation that is completely free of singularities that exist in the traditional formulation. The usual mathematical singularities that arise from using the fundamental solution in the conventional boundary integral method are removed by subtracting a related auxiliary flow field, $\boldsymbol{w}$, that can be constructed from one of many known fundamental solutions of the Stokes equation. This approach is exact and does not require the introduction of additional cutoff parameters. The numerical implementation of this boundary regularized integral equation formulation affords considerable savings in coding effort with improved numerical accuracy. The high accuracy of this formulation is retained even in problems where parts of the boundaries may almost be in contact.
\end{abstract}

%\pacs{47.15.G?, 47.11.Hj, 82.70.Dd, 83.50.Lh}% PACS, the Physics and Astronomy
                             % Classification Scheme.
%\keywords{Boundary regularized integral method, Stokes flow, fundamental solutions}%Use showkeys class option if keyword
                              %display desired
\maketitle

%==================================================
% -- SEC 1 --
\section{Introduction}

A common feature in the modelling of microfluidic design,\cite{Bazhlekov2004} the motion of particles with complex mixed stick-slip boundary conditions,\cite{Sun2013} the motility of biological cells \cite{Li2014} and the dynamics of cell-cell and cell-substrate interactions \cite{Bouffanais2010, Subramaniam2013} is the need to determine low-Reynolds-number Stokes flow in domains with boundaries having arbitrary shapes. In many applications, such boundaries may also deform as the quasi-static flow progresses, for instance, in modelling mobile deformable droplets or biological cells.\cite{Chan2011} Furthermore, such problems may require precision in resolution over very different characteristic length scales, for example, from micrometers for cell or droplet dimensions to nanometers for the deformation and spacing between interacting surfaces.\cite{Chan2011}

It is evident that modelling of the above problems using numerical methods based on the discretization of the 3D spatial domain will encounter a number of challenges. These include the need to insert and delete grid points when the boundaries deform and multi-grid methods might be needed to give the desired resolution and precision over different length scales. When high accuracy in locating boundaries is required, the 3D domain based discretisation algorithms can become impractical.

On the other hand, an approach based on the boundary integral formulation has a number of advantages for such problems. The most obvious is the reduction in dimension from 3D to 2D as the focus is entirely on the boundaries. Thus precision tracking of their deformation or motion becomes easier. Although the boundary integral formulation will give rise to a full coefficient matrix, the advent of the fast multipole method \cite{Liu2009} has reduced the computational cost to a very competitive $O(N \log N)$ level.

However, the conventional boundary integral formulation of Stokes flow problems involves singularities in the kernels that originate from the use of fundamental solutions of the Stokes equation.\cite{Becker92} Although such singularities are integrable, they do require careful analysis and additional coding effort in numerical implementation.\cite{Telles1987, Gaul2003} Furthermore, when two boundaries are close together, the singular behaviour on one surface can adversely affect the integral taken over the other nearby surface even in cases in which the field quantities are expected to be bounded. Also, the singular behavior in the flow domain near boundaries is often more difficult to deal with than the singularities on the boundaries.

Recently we have developed a reformulation of the boundary integral equations for the potential problem,\cite{Sun2014} the Helmholtz equation, the Stokes and linear elasticity problems in which the traditional singularities can be eliminated analytically.\cite{Klaseboer2012} This means that none of the above mentioned issues associated with the integrable singularities in the traditional boundary integral approach will arise and the integrals can be evaluated using any convenient quadrature method. Furthermore, the term involving the solid angle in the conventional boundary integral equation has also been eliminated, thus avoiding the need to calculate the solid angle at each node that is a complex function of local geometry.\cite{Mantic93}

In this paper, we show that corresponding to each of the many known fundamental solutions of the single-phase Stokes equation, a different regularized boundary integral equation for the Stokes problem can be derived. The regularization process results in numerically robust equations and this affords a choice of different numerically equivalent approaches for different applications.

In the next section, we give the general formulation of the boundary regularized integral equation formulation (BRIEF) for Stokes flow that involves finding an auxiliary flow field $\boldsymbol{w}$ that is used to remove all singularities analytically. In Sec. III, we give a new derivation of an earlier result using a linear flow field\cite{Klaseboer2012} that is a special case of a third order tensor fundamental solution to serve as a template on how to derive boundary regularized integral equations. In Sec. IV, we show how to construct $\boldsymbol{w}$ for other fundamental solutions that are also third order tensors and in Sec. V, we consider the construction of $\boldsymbol{w}$ from fundamental solutions that are second order tensors. For easy reference, all key results are summarized in Tables \ref{Tbl:M} and \ref{Tbl:D}. Numerical examples are given in Sec. VI to demonstrate the advantages of the BRIEF of Stokes flow. The structure of the algebraic system that arises from our approach is discussed in the Appendix.

%==================================================
% -- SEC 2 --
\section{Boundary regularized integral equation formulation (BRIEF) of Stokes flow}
The Stokes equation for low-Reynolds-number flow in a Newtonian fluid with dynamic shear viscosity, $\mu$: $-\nabla p + \mu \nabla^2 \boldsymbol{u} = 0$ can be written in Cartesian tensor notation  with the summation convention of repeated indices as
\begin{align}\label{eq:stokesflow1}
-\frac{\p{p}}{\p{x_i}}+\mu \frac{\p^2{u_{i}}}{\p{x_{k}}\p{x_{k}}} = 0
\end{align}
and the incompressibility condition: $\nabla \cdot \boldsymbol{u} = 0$ is
\begin{align}\label{eq:stokesflow2}
\frac{\p{u_{i}}}{\p{x_{i}}} = 0.
\end{align}
At the field point, $\boldsymbol{x}$ (with components $x_i$) where the pressure is $p(\boldsymbol{x})$ and the velocity field is $\boldsymbol{u}(\boldsymbol{x})$ (with components $u_i$), the stress tensor $\sigma_{ik}$ is given by
\begin{align}\label{eq:stokesstr}
\sigma_{ik} = -p \delta_{ik}+\mu\left[\frac{\p{u_i}}{\p{x_k}}+\frac{\p{u_k}}{\p{x_i}}\right],
\end{align}
in which $\delta_{ik}$ is the Kronecker delta function.

In the classic boundary integral formulation, the Lorentz reciprocal theorem~\citep{Lorentz1907} is used to give the following integral equation evaluated on the bounding surface(s), $S$, of the flow domain relating the velocity and the stress tensor on the boundary \citep {Pozrikidis92}
\begin{align}\label{eq:stkflbie}
c_0u^{0}_{j}+\int_{S} u_i{T}_{ijk}n_{k}\text{ d}S(\boldsymbol{x}) = \frac{1}{\mu}\int_{S}{\sigma}_{ik} n_{k}{U}_{ij} \text{ d}S(\boldsymbol{x}) \equiv \frac{1}{\mu}\int_{S}f_{i}{U}_{ij}\text{ d}S(\boldsymbol{x}).
\end{align}
In Eq. (\ref{eq:stkflbie}), $c_0$ is related to the solid angle at $\boldsymbol{x}_0$ that is located on the surface $S$. The component of the velocity at $\boldsymbol{x}_0$ in the $j$-th direction is $u_j^0\equiv u_j(\boldsymbol{x}_0)$, the component of the velocity at $\boldsymbol{x}$ in the $i$-th direction is $u_i\equiv u_i(\boldsymbol{x})$, and $n_{k}$ is the $k$-th component of the unit normal vector at position $\boldsymbol{x}$ on the surface, pointing out of the fluid domain. The $i$-th component of the traction vector $\boldsymbol{f}$, is defined as $f_{i} \equiv {\sigma}_{ik} n_k$. The kernels ${U}_{ij}$ and ${T}_{ijk}$ are the fundamental solutions for the 3D Stokes equation under a point force $\boldsymbol{g}$: $-\nabla p + \mu \nabla^2 \boldsymbol{u} +\boldsymbol{g} \delta(\boldsymbol{x} - \boldsymbol{x}_0) = \boldsymbol{0}$,~\citep {Pozrikidis92}
\begin{align}\label{eq:Ustokes}
{U}_{ij}(\boldsymbol{x},\boldsymbol{x}_0) = \frac{{\delta}_{ij}}{r}+\frac{\skew3\hat{x}_{i}\skew3\hat{x}_{j}}{r^3},
\end{align}
\begin{align}\label{eq:Tstokes}
{T}_{ijk}(\boldsymbol{x},\boldsymbol{x}_0) = -6\frac{\skew3\hat{x}_i\skew3\hat{x}_j\skew3\hat{x}_k}{r^5},
\end{align}
where $\skew3\hat{x}_i$ is the $i$-th component of $\skew3\hat{\boldsymbol{x}} \equiv \boldsymbol{x}-\boldsymbol{x}_0$, $r \equiv |\skew3\hat{\boldsymbol{x}}| \equiv  |\boldsymbol{x}-\boldsymbol{x}_0|$ (see Fig.~\ref{fig:x_d}), with corresponding velocity and traction fields $u_i= (1/8\pi\mu)U_{ij}g_j$ and $\sigma_{ik}=(1/8\pi)T_{ijk} g_j$. The kernels ${U}_{ij}$ and ${T}_{ijk}$ diverge as $1/r$ and $1/r^2$ respectively as $\boldsymbol{x}\rightarrow\boldsymbol{x}_0$, but the integrals over these singularities in Eq. (\ref{eq:stkflbie}) are finite even though their numerical evaluation requires careful treatment.\cite{Gaul2003} One method to deal with the singularities is to use a nonsingular contour-integral representation of the surface integrals, but the resulting computational performance is about an order of magnitude slower than the conventional approach.\cite{Bazhlekov2004} Another way to circumvent the divergence as $\boldsymbol{x}\rightarrow\boldsymbol{x}_0$ is to replace the Dirac $\delta$-function by spreading the applied force over a small ball of radius $\epsilon$ centered at $\boldsymbol{x}_0$.\cite{Cortez2001} The new length scale, $\epsilon$ has to be chosen carefully to give convergence without affecting the fidelity of solution since the fundamental solutions will also be modified. A ``near-singularity" subtraction method has also been proposed to handle the singularities in Eq. (\ref{eq:stkflbie}), but the method cannot completely eliminate the unbounded behavior of the double-layer integrand.\cite{Zinchenko1997}

% -- FIGURE 1 --
{
{\begin{figure}
\centering
\includegraphics[width=0.75\textwidth] {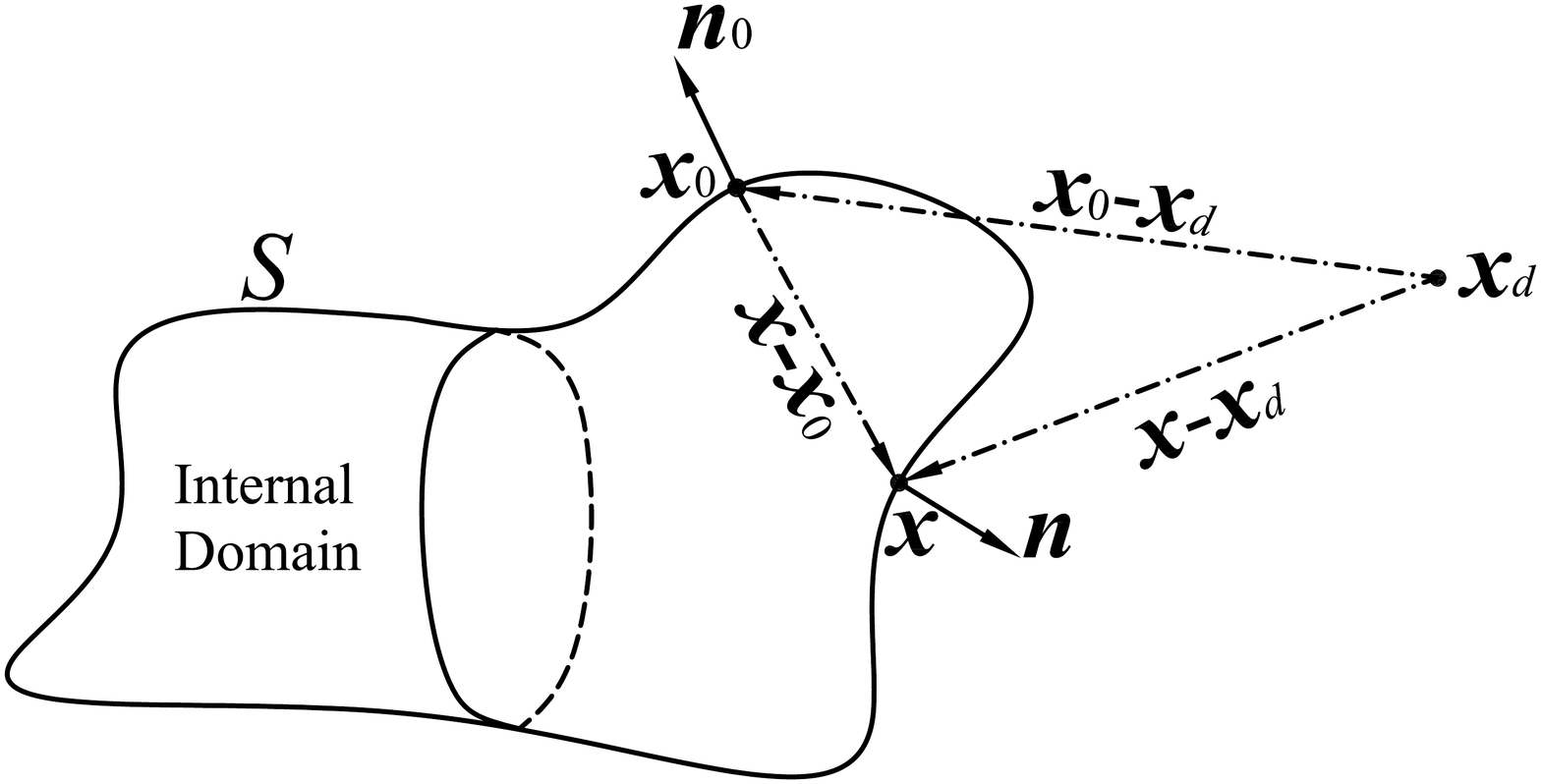}
\caption{The 3D internal domain defined by the closed surface $S$ showing the observation point $\boldsymbol{x_0}$ with outward normal $\boldsymbol{n}_0$, the integration point $\boldsymbol{x}$ with outward normal $\boldsymbol{n}$ and the general location of a point $\boldsymbol{x}_{d}$ outside the domain. Illustrated also are the relative position vectors: $\boldsymbol{x} - \boldsymbol{x}_{0} \equiv \skew3\hat{\boldsymbol{x}}\equiv\skew3\hat{x}_{i} = x_{i}-x_{i}^{0}$, $\boldsymbol{x} - \boldsymbol{x}_{d} \equiv \skew3\hat{\boldsymbol{x}}_{d}\equiv\skew3\hat{x}^{d}_{i} = x_{i}-x_{i}^{d}$ and $\boldsymbol{x}_{0} - \boldsymbol{x}_{d} \equiv \skew3\hat{\boldsymbol{a}}_{d}\equiv\skew3\hat{a}^{d}_{i} = x^{0}_{i}-x_{i}^{d}$. The magnitudes of these vectors are $r=|\skew3\hat{\boldsymbol{x}}|$, $r_{d}=|\skew3\hat{\boldsymbol{x}}_{d}|$ and $a=|\skew3\hat{\boldsymbol{a}}_{d}|$, with $b \equiv \skew3\hat{a}^{d}_{l}n_{l}^{0}\neq 0$.}\label{fig:x_d}
\end{figure}}
}

In earlier work,\cite{Klaseboer2012} we have shown that all the singularities associated with the kernels can be removed analytically, thus obviating the need to alter the nature of the fundamental solutions or to develop special integration algorithms to deal with the singularities. The way to achieve this is to construct an auxiliary known flow field, $\boldsymbol{w}(\boldsymbol{x})$ (with components $w_i$), for a given $\boldsymbol{x}_0$ in Eq. (\ref{eq:stkflbie}), that also satisfies the governing equations for the Stokes flow in Eqs. (\ref{eq:stokesflow1}) and (\ref{eq:stokesflow2}). Thus the field $\boldsymbol{w}$ satisfies
\begin{align}\label{eq:stkgnrlw1}
-\frac{\p{\mathcal{P}}}{\p{x_i}}+\mu \frac{\p^2{w_{i}}}{\p{x_{k} \p{x_{k} }}} = 0
\end{align}
and the compressibility condition
\begin{align}\label{eq:stkgnrlw2}
\frac{\p{w_{i}}}{\p{x_{i}}} = 0
\end{align}
in which $\mathcal{P}$ is the pressure corresponding to flow field $w_i$. The associated stress tensor ${\Sigma}_{ik}$ of this flow field is
\begin{align}\label{eq:stkgnrlwstr}
{\Sigma}_{ik} = -\mathcal{P}{\delta}_{ik}+\mu\left[\frac{\p{w_i}}{\p{x_k}}+\frac{\p{w_k}}{\p{x_i}}\right].
\end{align}
We further require $w_{i}$ to obey the following two constraints:
\begin{subequations}
\label{eq:nsbimcndtn}
\begin{align}
  \label{eq:nsbimcndtn_a}
  &w_{i}(\boldsymbol{x}_0)\equiv w^{0}_{i}=u^{0}_{i}\equiv u_{i}(\boldsymbol{x}_0), \\
  \label{eq:nsbimcndtn_b}
  &\mathcal{F}_{i}(\boldsymbol{x}_0)\equiv \mathcal{F}^{0}_{i} = f^{0}_{i}\equiv f_{i}(\boldsymbol{x}_0),
\end{align}
\end{subequations}
where $\mathcal{F}_{i} \equiv {\Sigma}_{ik}n_{k}$ is the traction vector of the flow field $w_{i}$.

Since $w_i$ satisfies the governing equations for incompressible Stokes flow, Eqs. (\ref{eq:stkgnrlw1}) and (\ref{eq:stkgnrlw2}), it also satisfies the boundary integral equation similar to Eq. (\ref{eq:stkflbie})
\begin{align}\label{eq:wbie}
c_0w^{0}_{j}+\int_{S} w_i{T}_{ijk}n_{k}\text{ d}S =  \frac{1}{\mu}\int_{S}\mathcal{F}_{i}{U}_{ij}\text{ d}S.
\end{align}
Subtracting Eq. (\ref{eq:wbie}) from Eq. (\ref{eq:stkflbie}), we get
\begin{align}\label{eq:stkflnsbie}
\int_{S} (u_i-w_{i}){T}_{ijk}n_{k}\text{ d}S = \frac{1}{\mu}\int_{S}(f_{i}- \mathcal{F}_{i}){U}_{ij}\text{ d}S.
\end{align}
This is the key general result of the boundary regularized integral equation formulation (BRIEF) for Stokes flow that relates the traction, $f_i$, and the surface velocity, $u_i$. The major advantages are that the term containing the solid angle $c_{0}$ is no longer present and that the integrands in Eq. (\ref{eq:stkflnsbie}) are regular over the entire integration surface $S$.\cite{Klaseboer2012} As a consequence, the surface integrals can be evaluated using any convenient quadrature method. The remaining task is to construct the flow field $w_{i}$ that will satisfy the two conditions given by Eq. (\ref{eq:nsbimcndtn}). Then Eq. (\ref{eq:stkflnsbie}) can be solved when either the traction, $f_i$, or the surface velocity, $u_i$, or a relation between $f_i$ and $u_i$ is specified from the prescribed boundary conditions of the problem.

%==================================================
% -- SEC 3 --
\section{BRIEF of Stokes flow using a linear flow field}
The simplest flow field $w_{i}$ that can be constructed to satisfy Eq. (\ref{eq:nsbimcndtn}) is the linear solution
\begin{align} \label{eq:linearW}
\boldsymbol{w} = \boldsymbol{u}_{0} + \frac{1}{\mu} {M} \bcdot (\boldsymbol{x} - \boldsymbol{x}_{0}) \equiv \boldsymbol{u}_{0} + \frac{1}{\mu} {M} \bcdot \skew3\hat{\boldsymbol{x}},
\end{align}
with an appropriate choice for the constant matrix ${M}$. The explicit form of ${M}$ for this case has been given earlier without proof.\cite{Klaseboer2012} Here we give a general derivation of this result that will provide guidance on how to construct other forms of $w_{i}$ using different fundamental solutions of the Stokes equation.

The linear flow field in Eq. (\ref{eq:linearW}) can be written as that due to a third order tensor
\begin{align}\label{eq:lnrw}
w_{i} = u_{i}^{0}+\frac{1}{\mu} {M}_{ik}\skew3\hat{x}_{k} = u_{i}^{0}+\frac{1}{\mu}{M}_{jk}\skew3\hat{x}_{k}{\delta}_{ij}.
\end{align}
We can see immediately that $w_{i}$ now satisfies Eq. (\ref{eq:stkgnrlw1}) with $\mathcal{P}=0$, and Eq. (\ref{eq:stkgnrlw2}) gives
\begin{align}\label{eq:lnrMtr}
%\frac{\p{w_{i}}}{\p{x_{i}}}=0 \Rightarrow {M}_{ll} \equiv \text{Tr}({M}) = 0.
{M}_{ll} \equiv \text{Tr}({M}) = 0.
\end{align}
The stress tensor, ${\Sigma}_{il}$, and traction, $\mathcal{F}_i$ , that correspond to $w_{i}$ are
\begin{align}\label{eq:lnrwtrF}
&{\Sigma}_{il}=\mu\left[\frac{\p{w_{i}}}{\p{x_{l}}} + \frac{\p{w_{l}}}{\p{x_{i}}}\right] = {M}_{jk}{\delta}_{kl}{\delta}_{ij} + {M}_{jk}{\delta}_{ki}{\delta}_{lj} = {M}_{il} + {M}_{li}, \nonumber\\
&\mathcal{F}_i \equiv {\Sigma}_{il}n_{l} = \left({M}_{il} + {M}_{li}\right)n_{l}.
\end{align}
Now ${M}$ must be chosen to satisfy: $\text{Tr}({M}) = 0$, and the traction, $\mathcal{F}_i$, must, according to Eq. (\ref{eq:nsbimcndtn_b}), be equal to $f^{0}_{i}$ at $\boldsymbol{x} = \boldsymbol{x}_0$, that is,
\begin{align}\label{eq:lnrwtr0}
\mathcal{F}_i^{0} \equiv \left[{M}_{il} + {M}_{li}\right]n^{0}_{l}=f^{0}_{i}.
\end{align}

Noting that the only vectors at our disposal are $\boldsymbol{f}_{0}$ and $\boldsymbol{n}_{0}$, then owing to the linear nature of the problem, ${M}_{jk}$ can only be a linear combination of the tensors: ${\delta}_{jk}$, $f_{j}^{0}n_{k}^{0}$, $n_{j}^{0}f_{k}^{0}$ and $n_{j}^{0}n_{k}^{0}$. One can see immediately that ${M}_{jk}$ should contain $f^{0}_{j}n^{0}_{k}$, since $M_{jk}{\delta}_{kl}{\delta}_{ij}n^{0}_{l}=(f^{0}_{j}n^{0}_{k}){\delta}_{kl}{\delta}_{ij}n^{0}_{l} =f_{i}^{0}$. However, terms with $f^{0}_{l}n^{0}_{l}$ will also appear due to the term $M_{jk}{\delta}_{ki}{\delta}_{lj}$ that gives $M_{jk}{\delta}_{ki}{\delta}_{lj}n^{0}_l=(f^{0}_{j}n^{0}_{k}){\delta}_{ki}{\delta}_{lj}n^{0}_{l} =f^{0}_{j}n^{0}_{i}n^{0}_{j}=(f^{0}_{l}n^{0}_{l})n^{0}_{i}$. Thus the most general form of ${M}_{jk}$ is
\begin{align}\label{eq:lnrMdf}
{M}_{jk} = f^{0}_{j}n^{0}_{k} + f^{0}_{l}n^{0}_{l}\left[c_{1}{\delta}_{jk}+c_{2}n^{0}_{j}n^{0}_{k}\right],
\end{align}
where $c_{1}$ and $c_{2}$ are constants and $c_{1}{\delta}_{jk}+c_{2}n^{0}_{j}n^{0}_{k}$ is the most general tensor, independent of $\boldsymbol{f}_{0}$ since it already appears in $f^{0}_{l}n^{0}_{l}$, that can be constructed from $n^{0}_{j}n^{0}_{k}$ (see \citet{Landau1966}). Combining Eqs. (\ref{eq:lnrMtr}), (\ref{eq:lnrwtr0}) and (\ref{eq:lnrMdf}), we have
\begin{align}
2c_{1}+2c_{2}+1 & =0,\nonumber\\
1+3c_{1}+c_{2} & =0,
\end{align}
with solution $c_{1}=c_{2}=-\frac{1}{4}$. Thus the desired form for ${M}_{jk}$ is
\begin{align}\label{eq:lnrMexprss}
{M}_{jk} = f^{0}_{j}n^{0}_{k} - \frac{1}{4} f^{0}_{l}n^{0}_{l}\left[{\delta}_{jk}+n^{0}_{j}n^{0}_{k}\right],
\end{align}
a result that was given earlier.\cite{Klaseboer2012} The linear solution for $\boldsymbol{w}$ can be used for both interior problems or exterior problems in an infinite domain, though not for semi-infinite domains.

%==================================================
% -- SEC 4 --
\section{BRIEF of Stokes flow using a third order tensor fundamental solution}
We have seen in the preceding section that the linear flow field $w_i$ considered in Eq. (\ref{eq:lnrw}) can be written in a more general form as a third order tensor
\begin{align} \label{eq:wQ}
w_{i} = u_{i}^{0} + \frac{1}{\mu} {M}_{jk} {Q}_{ijk},
\end{align}
in which ${M}_{jk}$ is a constant matrix and ${Q}_{ijk}$ is a third order tensor function of $\boldsymbol{x}$ that can be one of the fundamental solutions of Stokes flow.\cite{Blake1974, Chwang1975, Pozrikidis2011} The various known forms of ${Q}_{ijk}$ together with the corresponding pressure field $\mathcal{P}$ and coefficient matrix ${M}_{jk}$ are summarised in Table~\ref{Tbl:M}.

%\clearpage
%\newpage
% -- TABLE 1 --
\begin{table*}[htp]\centering
    %\captionsetup{justification=raggedright, singlelinecheck=false}
    \caption{The auxiliary flow field $w_{i} = u_{i}^{0} + \frac{1}{\mu} {M}_{jk} {Q}_{ijk}$ and the corresponding pressure, $\mathcal{P}$, constructed using fundamental solutions, ${Q}_{ijk}$, of the Stokes equation that are third order tensors. The domain of applicability are for interior problems bounded by a closed surface, exterior problems in an infinite domain or both. All other symbols are defined in the text or in Fig. 1.}\label{Tbl:M}
    \ra{1.3}
    \begin{ruledtabular}
    \begin{tabular}{p{8cm} l r r}
        \toprule
        Solution name & Pressure, $\mathcal{P}$ & $x_{i}^{d}$ & Domain \\
        \hline\hline \noalign{\smallskip}\noalign{\smallskip}
        (I) \textbf{Simple linear solution}\citep{Klaseboer2012} & Constant & No & Both \\
        \noalign{\smallskip}\noalign{\smallskip}
        \multicolumn{4}{l} {\hspace{0.2cm} ${Q}_{ijk} = \skew3\hat{x}_{k}{\delta}_{ij} $}  \\ \noalign{\smallskip}\noalign{\smallskip}
        \multicolumn{4}{l} {\hspace{0.2cm} ${M}_{jk} = f^{0}_{j}n^{0}_{k} - \frac{1}{4} f^{0}_{l}n^{0}_{l}\left[{\delta}_{jk}+n^{0}_{j}n^{0}_{k}\right]$}\\ \noalign{\smallskip}\noalign{\smallskip}
        \hline \noalign{\smallskip}\noalign{\smallskip}
        (II) \textbf{Stresson}\citep{Chwang1975, Pozrikidis2011} & Constant & No & Interior \\ \noalign{\smallskip}\noalign{\smallskip}
        \multicolumn{4}{l} {\hspace{0.2cm} ${Q}_{ijk} = 4\skew3\hat{x}_{j}{\delta}_{ki} - \skew3\hat{x}_{k}{\delta}_{ij} - \skew3\hat{x}_{i}{\delta}_{jk} $} \\ \noalign{\smallskip}\noalign{\smallskip}
        \multicolumn{4}{l} {\hspace{0.2cm} ${M}_{jk} = \frac{1}{6}\left(f^{0}_{j}n^{0}_{k} + n^{0}_{j}f^{0}_{k}\right) + \frac{1}{3} f^{0}_{l}n^{0}_{l}\left[{\delta}_{jk}- 4 n^{0}_{j}n^{0}_{k}\right]$} \\ \noalign{\smallskip}\noalign{\smallskip}
        \hline \noalign{\smallskip}\noalign{\smallskip}
        (III) \textbf{Stokeson dipole}\citep{Chwang1975, Pozrikidis2011}  & Constant & No & Interior \\ \noalign{\smallskip}\noalign{\smallskip}
        \multicolumn{4}{l} {\hspace{0.2cm} ${Q}_{ijk} = 4\skew3\hat{x}_{j}{\delta}_{ki} - \skew3\hat{x}_{k}{\delta}_{ij} - \skew3\hat{x}_{i}{\delta}_{jk} $} \\ \noalign{\smallskip}\noalign{\smallskip}
        \multicolumn{4}{l} {\hspace{0.2cm} ${M}_{jk} = \frac{1}{3}\left(n^{0}_{j}f^{0}_{k}\right) + \frac{1}{48} f^{0}_{l}n^{0}_{l}\left[{\delta}_{jk}- 19 n^{0}_{j}n^{0}_{k}\right]$ }\\ \noalign{\smallskip}\noalign{\smallskip}
        \hline \noalign{\smallskip}\noalign{\smallskip}
        (IV) \textbf{Source potential}\citep{Pozrikidis2011} & Constant & Yes & Both \\ \noalign{\smallskip}\noalign{\smallskip}
        \multicolumn{4}{l} {\hspace{0.2cm} ${Q}_{ijk} = \left(\frac{\skew3\hat{x}^{d}_{k}}{r_{d}^{3}} - \frac{\skew3\hat{a}^{d}_{k}}{a^{3}}\right){\delta}_{ij} $}  \\ \noalign{\smallskip}\noalign{\smallskip}
        \multicolumn{4}{l} {\hspace{0.2cm} ${M}_{jk} =  -\frac{a^{3}}{3b}\left(f^{0}_{j}\skew3\hat{a}^{d}_{k} - \skew3\hat{a}^{d}_{j}f^{0}_{k}\right)+ \frac{a^3}{2}\left[ \frac{a^2(f^{0}_{l}n^{0}_{l}) -2b(f^{0}_{l}\skew3\hat{a}^{d}_{l})} {a^{2}+b^{2}} \right] \left[ \frac{2}{3b} \left( n^{0}_{j}\skew3\hat{a}^{d}_{k} - \skew3\hat{a}^{d}_{j}n^{0}_{k} \right) + {\delta}_{jk} \right]$ }\\ \noalign{\smallskip}\noalign{\smallskip}
        \hline \noalign{\smallskip}\noalign{\smallskip}
        (V) \textbf{Stresslet}\citep{Chwang1975, Pozrikidis2011} & $2 {M}_{jk} \left(3\frac{\skew3\hat{x}^{d}_{j}\skew3\hat{x}^{d}_{k}}{r_{d}^{5}} - \frac{{\delta}_{jk}}{r_{d}^{3}} \right)$ & Yes & Both\\ \noalign{\smallskip}\noalign{\smallskip}
        \multicolumn{4}{l} {\hspace{0.2cm} ${Q}_{ijk} = \left( 3\frac{\skew3\hat{x}^{d}_{i}\skew3\hat{x}^{d}_{j}\skew3\hat{x}^{d}_{k}}{r_{d}^{5}} -\frac{\skew3\hat{x}^{d}_{i}{\delta}_{jk} }{r_{d}^{3}} \right) -\left( 3\frac{\skew3\hat{a}^{d}_{i}\skew3\hat{a}^{d}_{j}\skew3\hat{a}^{d}_{k}}{a^{5}} -\frac{\skew3\hat{a}^{d}_{i}{\delta}_{jk} }{a^{3}} \right) $}  \\ \noalign{\smallskip}\noalign{\smallskip}
        \multicolumn{4}{l} {\hspace{0.2cm} ${M}_{jk}=\frac{a}{12b^2}\skew3\hat{a}^{d}_{j}\skew3\hat{a}^{d}_{k} \left[-6b(f_{l}^{0}\skew3\hat{a}^{d}_{l}) + a^{2}(f_{l}^{0}n_{l}^{0}) \right] + \frac{a^3}{6b} (\skew3\hat{a}^{d}_{j}f_{k}^{0} + f_{j}^{0}\skew3\hat{a}^{d}_{k})$ }\\ \noalign{\smallskip}\noalign{\smallskip}
        \hline \noalign{\smallskip}\noalign{\smallskip}
        (VI) \textbf{Stokes doublet}\citep{Chwang1975, Pozrikidis2011} & $2 {M}_{jk} \left(3\frac{\skew3\hat{x}^{d}_{j}\skew3\hat{x}^{d}_{k}}{r_{d}^{5}} - \frac{{\delta}_{jk}}{r_{d}^{3}} \right)$ & Yes & Both \\ \noalign{\smallskip}\noalign{\smallskip}
        \multicolumn{4}{l} {\hspace{0.2cm} ${Q}_{ijk} =  \left[ \left( 3\frac{\skew3\hat{x}^{d}_{i}\skew3\hat{x}^{d}_{j}\skew3\hat{x}^{d}_{k}}{r_{d}^{5}} -\frac{\skew3\hat{x}^{d}_{i}{\delta}_{jk} }{r_{d}^{3}} \right) -\left( 3\frac{\skew3\hat{a}^{d}_{i}\skew3\hat{a}^{d}_{j}\skew3\hat{a}^{d}_{k}}{a^{5}} -\frac{\skew3\hat{a}^{d}_{i}{\delta}_{jk} }{a^{3}} \right)\right] + \left[\left(\frac{\skew3\hat{x}^{d}_{k}{\delta}_{ij} - \skew3\hat{x}^{d}_{j}{\delta}_{ik} }{r_{d}^{3}} \right) -\left( \frac{\skew3\hat{a}^{d}_{k}{\delta}_{ij} - \skew3\hat{a}^{d}_{j}{\delta}_{ik}}{a^{3}}\right)\right] $} \\ \noalign{\smallskip}\noalign{\smallskip}
        \multicolumn{4}{l} {\hspace{0.2cm} ${M}_{jk}=\frac{a^5}{6b^2} (n_{j}^{0}f_{k}^{0}) + \frac{a^3}{b}\skew3\hat{a}^{d}_{j}\skew3\hat{a}^{d}_{k} \left(\frac{f_{l}^{0}n_{l}^{0}}{6b}-\frac{5}{12a^2}f_{l}^{0}\skew3\hat{a}^{d}_{l}\right)$ }\\ \noalign{\smallskip}\noalign{\smallskip}
        \bottomrule
    \end{tabular}
    \end{ruledtabular}
\end{table*}
%\clearpage

We now show how the constant matrix ${M}_{jk}$ can be determined by ensuring that the conditions in Eq. (\ref{eq:nsbimcndtn}) are satisfied. Consider as an example, the Stokes stresslet fundamental solution~\citep{Chwang1975, Pozrikidis2011} given by item (V) in Table~\ref{Tbl:M} for which $w_i$ is
\begin{align}\label{eq:wletstrss}
w_{i} = u_{i}^{0} + \frac{1}{\mu}{M}_{jk}  \left[ \left(3\frac{\skew3\hat{x}^{d}_{i}\skew3\hat{x}^{d}_{j}\skew3\hat{x}^{d}_{k}}{r_{d}^{5}}  - \frac{\skew3\hat{x}^{d}_{i}{\delta}_{jk} }{r_{d}^{3}} \right) -\left( 3\frac{\skew3\hat{a}^{d}_{i}\skew3\hat{a}^{d}_{j}\skew3\hat{a}^{d}_{k}}{a^{5}} -\frac{\skew3\hat{a}^{d}_{i}{\delta}_{jk} }{a^{3}} \right)\right],
\end{align}
with the corresponding pressure field
\begin{align}\label{eq:wletstrssp}
\mathcal{P} = 2 {M}_{jk} \left(3\frac{\skew3\hat{x}^{d}_{j}\skew3\hat{x}^{d}_{k}}{r_{d}^{5}} - \frac{{\delta}_{jk}}{r_{d}^{3}} \right).
\end{align}
In Eqs. (\ref{eq:wletstrss}) and (\ref{eq:wletstrssp}), $\boldsymbol{x}_{d}\equiv x_{i}^{d}$ is the source position of the stresslet that is located outside the flow domain (see Fig.~\ref{fig:x_d}), $\skew3\hat{\boldsymbol{x}}_{d}\equiv\skew3\hat{x}^{d}_{i} = x_{i}-x_{i}^{d}$, $r_{d}=|\skew3\hat{\boldsymbol{x}}_{d}|$, $\skew3\hat{\boldsymbol{a}}_{d}\equiv\skew3\hat{a}^{d}_{i} = x^{0}_{i}-x_{i}^{d}$, $a=|\skew3\hat{\boldsymbol{a}}_{d}|$. Obviously, the flow field given in Eqs. (\ref{eq:wletstrss}) and (\ref{eq:wletstrssp}) satisfies Eqs. (\ref{eq:stkgnrlw1}) and (\ref{eq:stkgnrlw2}) as well as the first requirement of Eq. (\ref{eq:nsbimcndtn_a}): $w^{0}_{i}=u^{0}_{i}$, as $\boldsymbol{x}\rightarrow\boldsymbol{x}_0$. Noting that a stresslet is symmetric, the matrix ${M}_{jk}$ in Eq. (\ref{eq:wletstrss}) must also be symmetric. Thus, we can write the traction that corresponds to $w_i$ as
\begin{align}
\mathcal{F}_{i} = \frac{6}{r_{d}^{5}} \left[{M}_{ll}(\skew3\hat{x}^{d}_{l}n_{l})\skew3\hat{x}^{d}_{i} + (\skew3\hat{x}^{d}_{j}{M}_{jk}n_{k}) \skew3\hat{x}^{d}_{i}
 + \skew3\hat{x}^{d}_{j}{M}_{ji}(\skew3\hat{x}^{d}_{l}n_{l}) - \frac{5}{r_{d}^{2}} (\skew3\hat{x}^{d}_{j}{M}_{jk}\skew3\hat{x}^{d}_{k}) (\skew3\hat{x}^{d}_{l}n_{l}) \skew3\hat{x}^{d}_{i} \right].
\end{align}
The condition in Eq. (\ref{eq:nsbimcndtn_b}) implies
\begin{align}\label{eq:mstksdf}
f_{i}^{0} = \frac{6}{a^{5}} &\left[{M}_{ll}(\skew3\hat{a}^{d}_{l}n_{l}^{0})\skew3\hat{a}^{d}_{i} + (\skew3\hat{a}^{d}_{j}{M}_{jk}n_{k}^{0}) \skew3\hat{a}^{d}_{i} + \skew3\hat{a}^{d}_{j}{M}_{ji}(\skew3\hat{a}^{d}_{l}n_{l}^{0}) - \frac{5}{a^{2}} (\skew3\hat{a}^{d}_{j}{M}_{jk}\skew3\hat{a}^{d}_{k}) (\skew3\hat{a}^{d}_{l}n_{l}^{0}) \skew3\hat{a}^{d}_{i} \right].
\end{align}
There are a few different choices for ${M}_{jk}$ that can satisfy the above constraint. The vectors at our disposal are $\boldsymbol{f}_{0}$, $\boldsymbol{n}_{0}$ and $\skew3\hat{\boldsymbol{a}}_{d}$, so to preserve the linear nature of ${M}_{jk}$ with respect to $f_{i}^{0}$, it should be a linear combination of ${\delta}_{jk}$, $f_{j}^{0}\skew3\hat{a}^{d}_{k}$, $\skew3\hat{a}^{d}_{j}f_{k}^{0}$, $f_{j}^{0}n_{k}^{0}$, $n_{j}^{0}f_{k}^{0}$, $n_{j}^{0}\skew3\hat{a}^{d}_{k}$, $\skew3\hat{a}^{d}_{j}n_{k}^{0}$, $\skew3\hat{a}^{d}_{j}\skew3\hat{a}^{d}_{k}$ and $n_{j}^{0}n_{k}^{0}$. One possible option for such a linear combination is
\begin{align}
{M}_{jk}=\frac{a}{12b^2}\skew3\hat{a}^{d}_{j}\skew3\hat{a}^{d}_{k} \left[-6b(f_{l}^{0}\skew3\hat{a}^{d}_{l}) + a^{2}(f_{l}^{0}n_{l}^{0}) \right] + \frac{a^3}{6b} (\skew3\hat{a}^{d}_{j}f_{k}^{0} + f_{j}^{0}\skew3\hat{a}^{d}_{k}),
\end{align}
in which $b=\skew3\hat{a}^{d}_{l}n_{l}^{0}\neq 0$. This is the result given in item (V) in Table~\ref{Tbl:M}.

Different formulations of the regular form of the boundary integral equation~(\ref{eq:stkflnsbie}) using other fundamental solutions, ${Q}_{ijk}$, of the Stokes equation that are third order tensors together with the corresponding pressure field, $\mathcal{P}$, and coefficient matrix, ${M}_{jk}$, are given in Table~\ref{Tbl:M}. They can be readily derived by following the steps outlined above.

%==================================================
% -- SEC 5 --
\section{BRIEF of Stokes flow using a second order tensor fundamental solution}
Fundamental solutions for Stokes flow that are second order tensor functions, ${S}_{ij}$, can also be used to construct the flow field, $w_i$ of the general form
\begin{align} \label{eq:wS}
w_{i} = u_{i}^{0} + \frac{1}{\mu} D_{j} {S}_{ij}
\end{align}
where the constant vector, $D_{j}$, is found by imposing the two conditions in Eq. (\ref{eq:nsbimcndtn}).

One example of this class of solutions is the source potential doublet,\cite{Chwang1975, Pozrikidis2011} see item (II) of Table \ref{Tbl:D}, that has zero pressure, $\mathcal{P}=0$. The flow field $w_{i}$ that can be constructed is
\begin{align}\label{eq:wspd}
w_{i}=u^{0}_{i} + \frac{1}{\mu} D_{j}\left\{\left[\frac{{\delta}_{ij}}{r^{3}_{d}} -\frac{3 \skew3\hat{x}_{i}^{d}\skew3\hat{x}_{j}^{d} }{r^{5}_{d}} \right] - \left[\frac{{\delta}_{ij}}{a^{3}} -\frac{3 \skew3\hat{a}^{d}_{i}\skew3\hat{a}^{d}_{j} }{a^{5}} \right] \right\}
\end{align}
that satisfies Eq. (\ref{eq:nsbimcndtn_a}). The corresponding traction is
\begin{align}\label{eq:trspd}
\mathcal{F}_{i} = -\frac{6}{r_{d}^{5}} \left[D_{i}(\skew3\hat{x}^{d}_{l}n_{l}) + \skew3\hat{x}^{d}_{i}(n_{l}D_{l}) + n_{i}(D_{l}\skew3\hat{x}^{d}_{l}) - \frac{5}{r_{d}^{2}} \skew3\hat{x}^{d}_{i} (D_{l}\skew3\hat{x}^{d}_{l})(\skew3\hat{x}^{d}_{k}n_{k})  \right].
\end{align}
The requirement of Eq. (\ref{eq:nsbimcndtn_b}) on the traction then implies
\begin{align}
f^{0}_{i} = -\frac{6}{a^{5}} \left[b D_{i} + \skew3\hat{a}^{d}_{i}(n^{0}_{l}D_{l}) + n^{0}_{i}(D_{l}\skew3\hat{a}^{d}_{l}) - \frac{5b}{a^{2}} \skew3\hat{a}^{d}_{i} (D_{l}\skew3\hat{a}^{d}_{l})  \right],
\end{align}
in which $b=\skew3\hat{a}^{d}_{l}n_{l}^{0}\neq 0$. This gives
\begin{align}
D_{j} = -\frac{a^{5}f_{j}^{0}}{6b} & + \frac{a^{5}n_{j}^{0}}{6b} \left[ \frac{a^2(f^{0}_{l}n^{0}_{l}) - 2b(f^{0}_{l}\skew3\hat{a}^{d}_{l})} {a^2 + b^2} \right] \nonumber \\
& + \frac{\skew3\hat{a}^{d}_{j}}{6b}\left[ \frac{-2 a^5 b (f^{0}_{l}n^{0}_{l}) + (a^5+5a^3b^2)(f^{0}_{l}\skew3\hat{a}^{d}_{l})}{a^2 + b^2}\right].
\end{align}

Table \ref{Tbl:D} also contains the result for $w_i$ obtained using a Stokeson.

% -- TABLE 2 --
\begin{table*}\centering
    %\captionsetup{justification=raggedright, singlelinecheck=false}
    \caption{The auxiliary flow field $w_{i} = u_{i}^{0} + \frac{1}{\mu} D_{j} {S}_{ij}$ and the corresponding pressure, $\mathcal{P}$, constructed using fundamental solutions, ${S}_{ij}$, of the Stokes equation that are second order tensors. All other symbols are defined in the text or in Fig. 1.}\label{Tbl:D}
    \ra{1.3}
    \begin{ruledtabular}
    \begin{tabular}{p{8cm} l r r}
        \toprule
        Solution name & Pressure, $\mathcal{P}$ & $x_{i}^{d}$ & Domain \\
        \hline\hline \noalign{\smallskip}\noalign{\smallskip}
        (I) \textbf{Stokeson}\citep{Chwang1975, Pozrikidis2011}  & $10 D_{j} \skew3\hat{x}^{d}_{j}$ & Yes & Interior \\ \noalign{\smallskip}\noalign{\smallskip}
        \multicolumn{4}{l} {\hspace{0.2cm} ${S}_{ij} =  (2r_{d}^{2}{\delta}_{ij} - \skew3\hat{x}^{d}_{i}\skew3\hat{x}^{d}_{j}) - (2a^{2}{\delta}_{ij} - \skew3\hat{a}^{d}_{i}\skew3\hat{a}^{d}_{j}) $}  \\ \noalign{\smallskip}\noalign{\smallskip}
        \multicolumn{4}{l} {\hspace{0.2cm} $D_{j} = \left[ \frac{2f_{l}^{0}\skew3\hat{a}^{d}_{l}} {b(9b^2 - 6 a^2)} - \frac{f_{l}^{0}n_{l}^{0}} {6b^2 - 4 a^2} \right] [-2 b n_{j}^{0}  + \skew3\hat{a}^{d}_{j}] +\frac{1}{3b} [f_{j}^{0} - (f_{l}^{0}n_{l}^{0})n_{j}^{0}] $} \\
        \noalign{\smallskip}\noalign{\smallskip}
        \hline \noalign{\smallskip}\noalign{\smallskip}
        (II) \textbf{Source potential doublet}\citep{Chwang1975, Pozrikidis2011}  & Constant & Yes & Both \\ \noalign{\smallskip}\noalign{\smallskip}
        \multicolumn{4}{l} {\hspace{0.2cm} ${S}_{ij} =  \left[\frac{{\delta}_{ij}}{r^{3}_{d}} -\frac{3 \skew3\hat{x}_{i}^{d}\skew3\hat{x}_{j}^{d} }{r^{5}_{d}} \right] - \left[\frac{{\delta}_{ij}}{a^{3}} -\frac{3 \skew3\hat{a}^{d}_{i}\skew3\hat{a}^{d}_{j} }{a^{5}} \right] $} \\ \noalign{\smallskip}\noalign{\smallskip}
        \multicolumn{4}{l} {\hspace{0.2cm} $D_{j} = -\frac{a^{5}f_{j}^{0}}{6b}  + \frac{a^{5}n_{j}^{0}}{6b} \left[ \frac{a^2(f^{0}_{l}n^{0}_{l}) - 2b(f^{0}_{l}\skew3\hat{a}^{d}_{l})} {a^2 + b^2} \right] + \frac{\skew3\hat{a}^{d}_{j}}{6b}\left[ \frac{-2 a^5 b (f^{0}_{l}n^{0}_{l}) + (a^5+5a^3b^2)(f^{0}_{l}\skew3\hat{a}^{d}_{l})}{a^2 + b^2}\right]$ }\\ \noalign{\smallskip}\noalign{\smallskip}
        \bottomrule
    \end{tabular}
    \end{ruledtabular}
\end{table*}

%==================================================
% -- SEC 6 --
\section{Examples}
We now furnish examples to illustrate the implementation of the boundary regularized integral equation formulation (BRIEF) of single-phase Stokes flow problems and to demonstrate the many unique advantages of this approach. The key result of the BRIEF given by Eq.~(\ref{eq:stkflnsbie}) relates the surface velocity to the surface traction. Therefore we first demonstrate the solution of two problems that are defined either by specifying the velocity field or the surface traction. We also use this example to quantify the precision that can be gained by using quadratic instead of linear surface elements while keeping the number of unknowns or degrees of freedom consistent, and to show the numerical equivalence of using different fundamental solutions to regularize the integral equation.

A second example will be used to demonstrate that it is straightforward to calculate the velocity field accurately not only far from but also close to boundaries using the BRIEF because of the complete absence of singular behavior. This is a very distinct advantage over the conventional boundary integral method in which singularities of the kernel can adversely affect the accurate evaluation of the velocities at field points close to the surface.

Another advantage of the BRIEF is that extreme geometric aspect ratios in the boundaries do not degrade the numerical precision of the solution. We illustrate this by considering the case of two nearly touching spheres in an external flow field that also serves the purpose of illustrating how forces and torques can be computed accurately by the BRIEF. In this and subsequent examples, we also consider cases in which neither the surface velocity nor the surface traction are specified, but rather a relation between these two quantities obeying the Navier slip boundary condition is imposed.

We also consider examples of lubrication flow between closely spaced surfaces where the field quantities become unbounded in the limit of zero separation. These examples illustrate the fact that in absence of mathematical singularities, the BRIEF is better positioned to handle the unavoidable physical divergences.

Finally, we consider the calculation of forces and torques experienced by bodies of varying geometric aspect ratios with varying Navier slip boundary conditions imposed on the surfaces.

Where appropriate, we compare results with that obtained from the conventional boundary integral method (CBIM) in which the rigid body solution has been subtracted:\cite{Becker92}
\begin{align}\label{eq:stkcbim}
\int_{S} (u_i-u^{0}_{i}){T}_{ijk}n_{k}\text{ d}S(\boldsymbol{x}) = \frac{1}{\mu}\int_{S}f_{i}{U}_{ij}\text{ d}S(\boldsymbol{x}).
\end{align}

There is a general point to note in selecting the desired flow field $\boldsymbol{w}$ given in Tables \ref{Tbl:M} and \ref{Tbl:D} to construct the BRIEF. For external problems, it is also necessary to consider the integral over the surface at infinity, $S_\infty$, that grows at $r^2$ as $r \rightarrow \infty$. For the cases of $\boldsymbol{w}$ that have been identified as applicable to both internal and external problems under the Domain category in Tables \ref{Tbl:M} and \ref{Tbl:D}, the terms that involve $r_{d}$ decay faster than $r^{-2}$ and so such terms do not contribute to the integral over $S_\infty$. The integral on the right hand side of Eq. (\ref{eq:stkflnsbie}) also vanishes on $S_\infty$ because $U_{ij}$ vanishes as $r^{-1}$, $f_i$ as $r^{-2}$ and $\mathcal{F}_i$ as $r^{-3}$ as $r \rightarrow \infty$. Finally, the limiting form of the left hand side of Eq. (12) on $S_\infty$ will depend on the choice of the flow field $w_i$ constructed from the fundamental solutions. As an example, for the case of a stresslet (entry V in Table I), the left hand side of Eq. (\ref{eq:stkflnsbie}) on $S_\infty$ is
\begin{align}
8\pi u^{0}_{i} -\frac{8\pi}{\mu} \skew3\hat{a}^{d}_{i}\left[\frac{1}{3b} (f^{0}_{l}\skew3\hat{a}^{d}_{l})-\frac{a^2}{6b^2} (f^{0}_{l}n^{0}_{l})\right].
\end{align}
A notable exception is the linear solution for which the contribution on $S_\infty$ also vanishes due to cancellations owing to the symmetry of the angular integration when the domain is infinite. If this is not the case, for instance, in a semi-infinite domain, the linear solution cannot be used.

\subsection{Velocity or traction boundary conditions}

{  % -- Figure 2 a & b

{\begin{figure}[t]
\centering
\subfloat[]{ \includegraphics[width=0.47\textwidth] {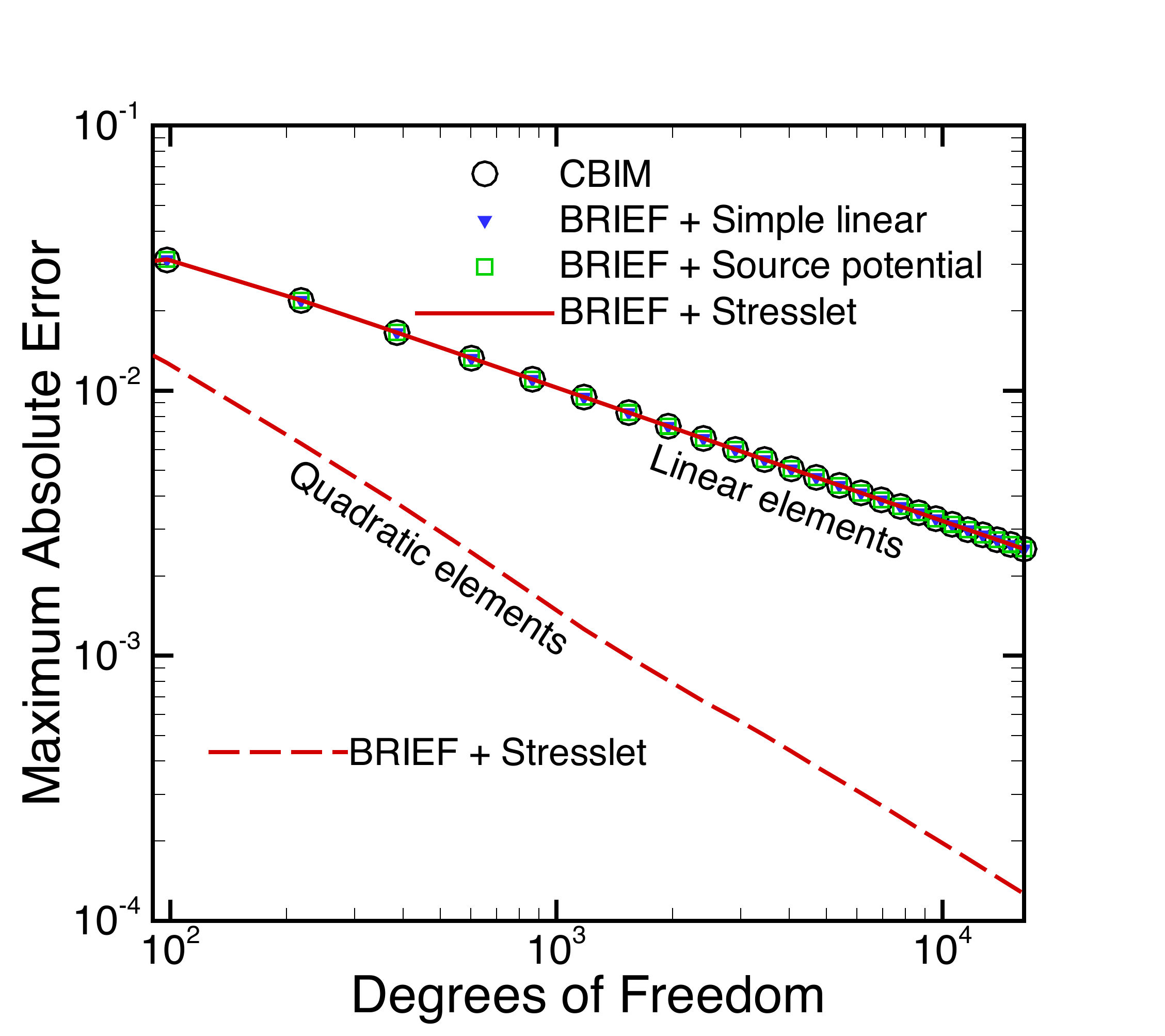} }
\subfloat[]{ \includegraphics[width=0.47\textwidth] {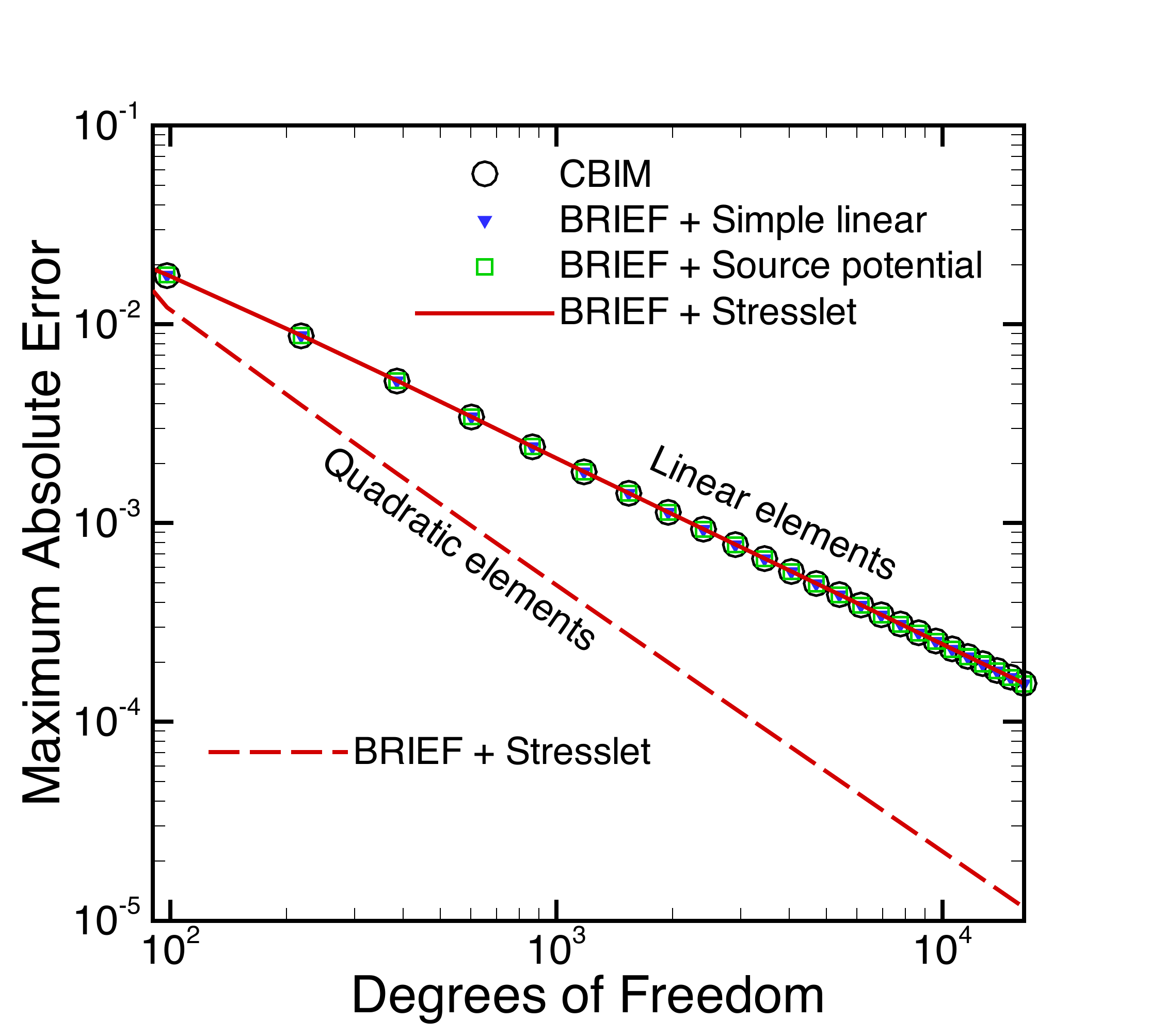} }
\caption{The maximum absolute error of the surface velocity on a sphere: (a) with the zero tangential stress boundary condition located in an uniform flow field and (b) in a quiescent fluid under a prescribed surface traction $\boldsymbol{f} = (3\mu U/R^2) z \boldsymbol{n}$.\cite{Pigeonneau2011} The BRIEF results, Eq. (\ref{eq:stkflnsbie}), obtained using a linear field, a source potential and a stresslet for the auxillary field $\boldsymbol{w}$ are compared to results from the conventional boundary integral method (CBIM) according to Eq. (\ref{eq:stkcbim})}\label{fig:2}
\end{figure}}
}

We consider two examples in which we determine the velocity on a boundary that is either a sphere with free slip or zero tangential stress boundary conditions placed in an external flow field, $U\boldsymbol{k}$, that is uniform at infinity, or on a boundary that is a sphere centered at the origin in a quiescent fluid and subjected to a surface traction $\boldsymbol{f} = (3\mu U/R^2) z \boldsymbol{n}$. In the former case, the velocity components parallel and perpendicular to the surface are given by~\cite{Lamb1932}: $u_\bot = U\cos\theta\{1-(R/r_c)\}$,  $u_{||} = U \sin\theta\{-1+\tfrac{1}{2}(R/r_c)\}$, where $r_c$ is the radial distance from the center of the sphere and $\theta$ is the angle between the vector $\boldsymbol{k}$ and the radial direction.

In these two examples, we compare in Fig.~\ref{fig:2}, the variation of the maximum absolute error in the surface velocity as a function of the number of degrees of freedom (or the number of unknowns) using linear and quadratic elements based on the same triangular surface mesh. With linear elements, we see that results from  the conventional boundary integral method (CBIM) and from the boundary regularized integral equation formulation (BRIEF) using a linear function, a source potential or a stresslet with $\boldsymbol{x}_d$ at the center of the sphere to remove the singularities, gave practically identical results. However, if quadratic elements are used, the magnitude of the maximum absolute error decreases by almost an order of magnitude, or equivalently, if quadratic elements are used instead of linear elements, the same precision can be attained be reducing the degree of freedom by about a factor of 10. Thus the combination of the BRIEF with quadratic elements offers  significant advantages in terms of precision and computational effort.

\subsection{Velocity field near boundaries}

The boundary regularized integral equation formulation (BRIEF) of Stokes flow can also be used to calculate accurately the velocity at field points close to a boundary by alleviating the loss of precision in the conventional boundary integral method (CBIM) due to the near singular behavior of the integrals. We achieve this by using a simple extension of the method developed for the solution of the Laplace equation\cite{Sun2014} to give a robust way to calculate the flow velocity by BRIEF anywhere within the flow domain using:
\begin{align}\label{eq:Bdmn}
u_{j}(\boldsymbol{x}_p) = w_{j}(\boldsymbol{x}_p) - \frac{1}{8\pi}\left[ \int_{S} (u_{i}-w_{i})({T}^{p}_{ijk}-{T}_{ijk})n_{k}\text{ d}S(\boldsymbol{x}) - \frac{1}{\mu} \int_{S}(f_{i}- \mathcal{F}_{i})({U}^{p}_{ij}-{U}_{ij})\text{ d}S(\boldsymbol{x}) \right].
\end{align}
Here $\boldsymbol{x}_p$ is a point in the flow domain for which the fluid velocity $u_{j}(\boldsymbol{x}_p)$ is to be calculated, $w_{j}(\boldsymbol{x}_p)$ is value of the constructed flow field at the same position, $U^p_{ij} = U_{ij}(\boldsymbol{x},\boldsymbol{x}_p)$, $T^p_{ijk} = T_{ijk}(\boldsymbol{x},\boldsymbol{x}_p)$, and $\boldsymbol{x}_0$ in $U_{ij}(\boldsymbol{x},\boldsymbol{x}_0)$, $T_{ijk}(\boldsymbol{x},\boldsymbol{x}_0)$ is taken to be the node on the boundary, $S$. When $\boldsymbol{x}_p$ is close to the surface, we choose $\boldsymbol{x}_0$ to be related to $\boldsymbol{x}_p$ by: $\boldsymbol{x}_p = \boldsymbol{x}_0 + \alpha \boldsymbol{n}_0$, where $\alpha$ is a small constant. With this choice, the near singular behavior of the term $({T}^{p}_{ijk}-{T}_{ijk})n_{k}$ is alleviated.

{  % -- Figure 3 a & b
{\begin{figure}[t]
\centering
\subfloat[]{ \includegraphics[width=0.47\textwidth] {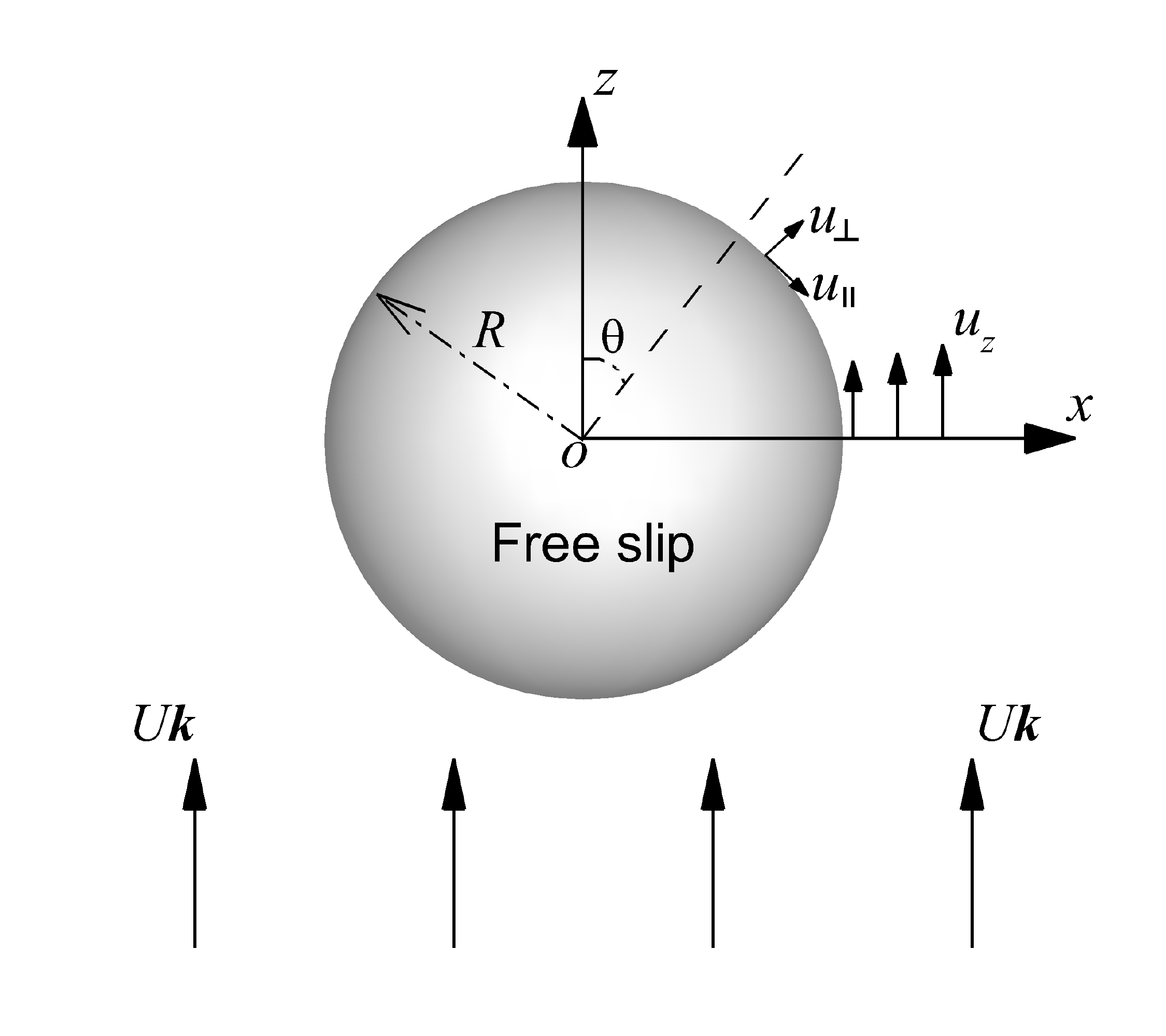} }
\subfloat[]{ \includegraphics[width=0.47\textwidth] {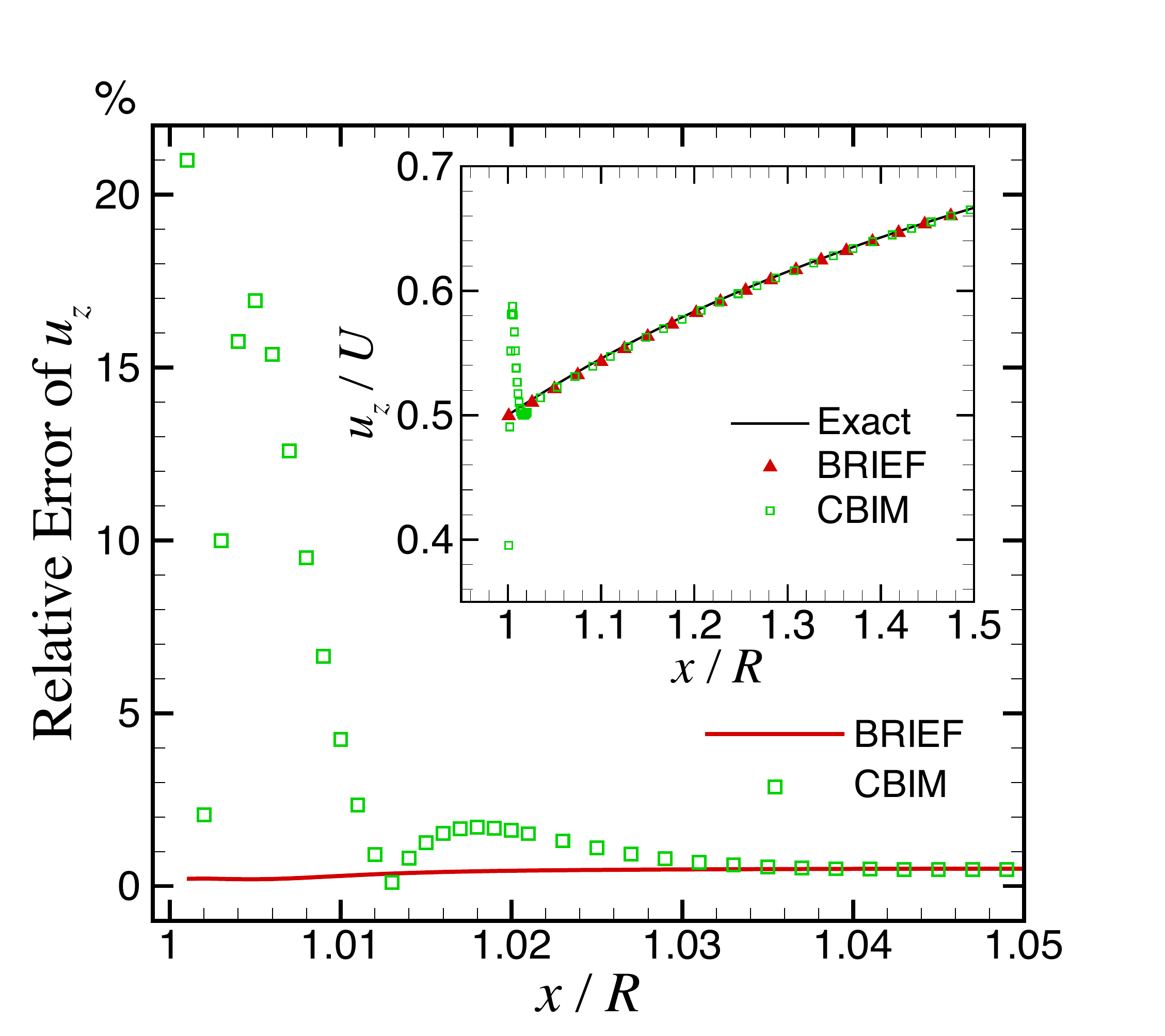} }
\caption{The absolute relative error of the $z$-component, $u_z$ of the velocity on the equatorial plane obtained by the boundary regularized integral equation formulation (BRIEF) of Stokes flow and by the conventional boundary integral method (CBIM) for a free slip sphere with the zero tangential stress boundary condition in a uniform external field $U\boldsymbol{k}$ using 2352 linear elements with 1178 nodes and the source potential for $\boldsymbol{w}$. Inset: The scaled velocity, $u_z/U$ as a function of position.}\label{fig:3}
\end{figure}}
}

To illustrate the accuracy that can be obtained by using the BRIEF, we consider a sphere of radius, $R$, with the free slip (zero tangential stress) boundary condition in a uniform flow field, $U\boldsymbol{k}$, at infinity and compare the velocity in the equatorial plane $(z=0)$ as a function of position with the known analytical result\cite{Lamb1932}
\begin{align}\label{eq:LambFSDmn}
u_{z} (z=0)= U\left(1 - \frac{R}{2x}\right).
\end{align}
From Fig.~\ref{fig:3}, we see that using the BRIEF plus the source potential (item IV in Table \ref{Tbl:M} with $\boldsymbol{x}_d$ at the center of the sphere) with 2352 linear trangular elements and 1178 nodes, the relative error is less than 1\% at all positions. In contrast, the relative error using the CBIM with the same surface mesh can exceed 20\% close to the sphere surface. This indicates that the BRIEF also alleviates the near singular behavior in the flow domain close to boundaries that is often more difficult to deal with than the singular behavior on the boundaries in CBIM.

\subsection{Nearly touching surfaces}
Another significant advantage of the BRIEF of Stokes flow is in cases in which boundaries are very close together whereby the problem has disparate but important characteristic length scales.  In the conventional boundary integral formulation (CBIM), the singular behavior of the kernel on one boundary will invariably have an adverse effect on the precision of integrals evaluated on the other nearby boundary even in instances in which the physical problem does not have any singular behavior. On the other hand, the complete absence of singular terms in the BRIEF for Stokes flow means that such numerical problems do not arise.

We illustrate this with the example of two nearly touching spheres of radius $R$ in a uniform flow field at infinity oriented either parallel or perpendicular to the line of centers. The sphere surfaces either have a free slip (zero tangential stress) condition or a no slip (zero velocity) condition. Results are given for the drag force, $\boldsymbol{F}$, on each sphere calculated by integrating the traction on the surface, $\bar{S}$, of that sphere
\begin{align}
\boldsymbol{F} = \int_{\bar{S}} \boldsymbol{f} \text{ d}S(\boldsymbol{x}),
\end{align}
and similarly for the torque, $\boldsymbol{N}$, about the center of the sphere at $\boldsymbol{x}_{c}$,
\begin{align}
\boldsymbol{N} = \int_{\bar{S}} (\boldsymbol{x} - \boldsymbol{x}_{c}) \times \boldsymbol{f} \text{ d}S(\boldsymbol{x}).
\end{align}
The force and torque are calculated as functions of the distance of closest approach, $h/R$, between the spheres. The results are obtained using the source potential (item IV in Table \ref{Tbl:M} with $\boldsymbol{x}_d$ at the center of the sphere on which $\boldsymbol{x}_0$ is located) solution in the BRIEF with 2352 linear elements and 1178 nodes on each sphere.

For two spheres aligned with their line of centers along the uniform flow field, we show in Fig. 4, the drag force as a function of sphere separation for the combinations of free slip boundary condition on both spheres, no slip boundary conditions on both spheres, no slip boundary condition on the upstream sphere with free slip boundary condition on the downstream sphere as well as for two spheres that obey the Navier slip boundary condition with a prescribed slip length, $s$
\begin{align}
\left\{\begin{aligned}&u_{n} = 0,\\
&u_{t1} = (s/\mu) f^{S}_{t1},\\
&u_{t2} = (s/\mu) f^{S}_{t2},
\end{aligned}\right . \quad s \ge 0.
\end{align}
Here $u_{n} $ is normal component and $u_{t1}$ and $u_{t2}$ are the tangential components of the fluid velocity on the surface of the sphere. The tangential components of the traction acting on the sphere are denoted by $f^{S}_{t1}$ and $f^{S}_{t2}$. The limit of the slip length $s = 0$, reduces to the familiar no-slip condition with $\boldsymbol{u} = 0$ on the surface. The limit $s \rightarrow \infty$ is the perfect free-slip or zero tangential stress case with the boundary conditions
$$
u_{n} = 0,\quad f^{S}_{t1} = 0,\quad f^{S}_{t2} = 0.
$$
In cases for which exact analytic solutions are available~\cite{Reed1974}, the agreement with BRIEF is excellent (see Fig 4).

In Fig. 5, we show corresponding results for the surface velocity vector field for the free slip/free slip and free slip/no slip combinations at a small distance of closest approach between the spheres: $h/R=0.01$ to illustrate the capabilities of the BRIEF. In Fig. 6, the traction field for all three combinations of free slip/free slip, free slip/no slip and no slip/no slip at the same separation are shown.

{  % -- Figure 4 a & b
{\begin{figure}[t]
\centering
\subfloat[]{ \includegraphics[width=0.47\textwidth] {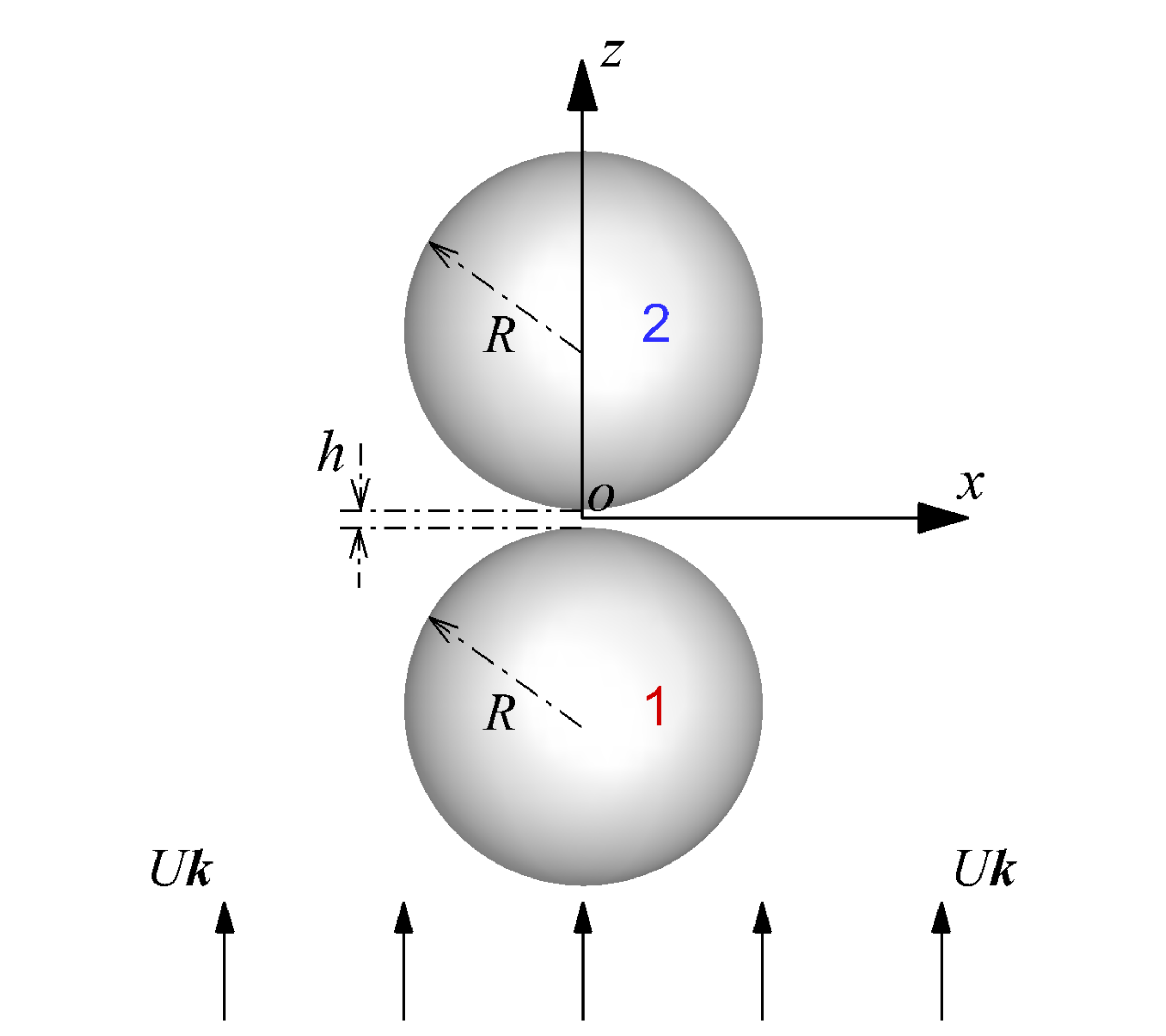} }
\subfloat[]{ \includegraphics[width=0.47\textwidth] {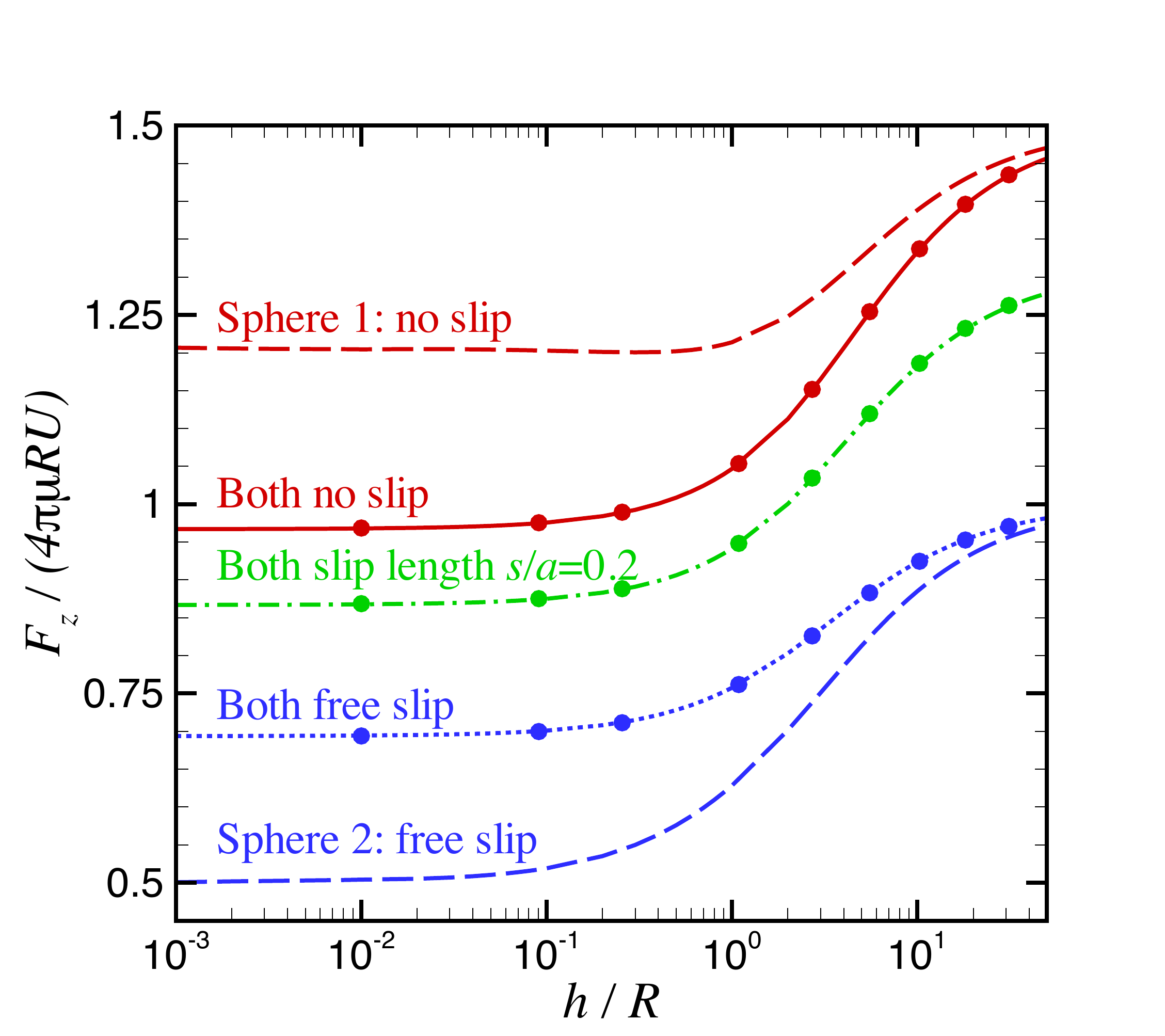} }
\caption{(a) The geometry of two spheres in an uniform flow along the line of centers. (b) Drag force in the $z$-direction, $F_z$, as a function of separation, $h/R$, for different combinations of no slip, finite slip length, $s/a = 0.2$ and free slip boundary conditions on the spheres compared with exact analytical solutions ($\bullet$) using bipolar coordinates\cite{Reed1974}. The BRIEF results (lines) with maximum relative error of less than 0.1\% are obtained using 2352 linear elements with 1178 nodes on each sphere and the source potential for $\boldsymbol{w}$.}
\end{figure}}
}

{  % -- Figure 5
{\begin{figure}[t]
\centering
\subfloat[]{ \includegraphics[width=0.31\textwidth] {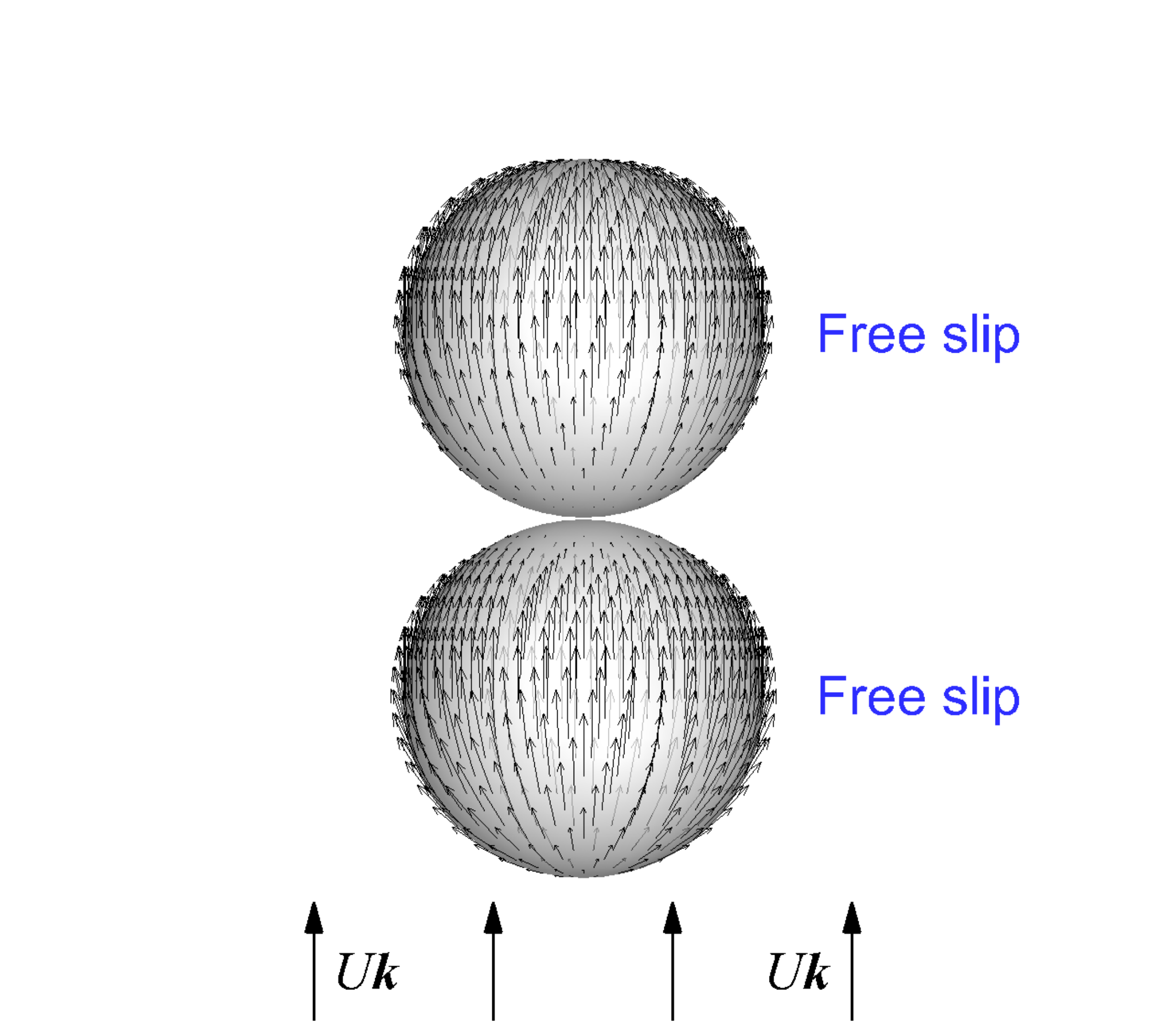} }
\subfloat[]{ \includegraphics[width=0.31\textwidth] {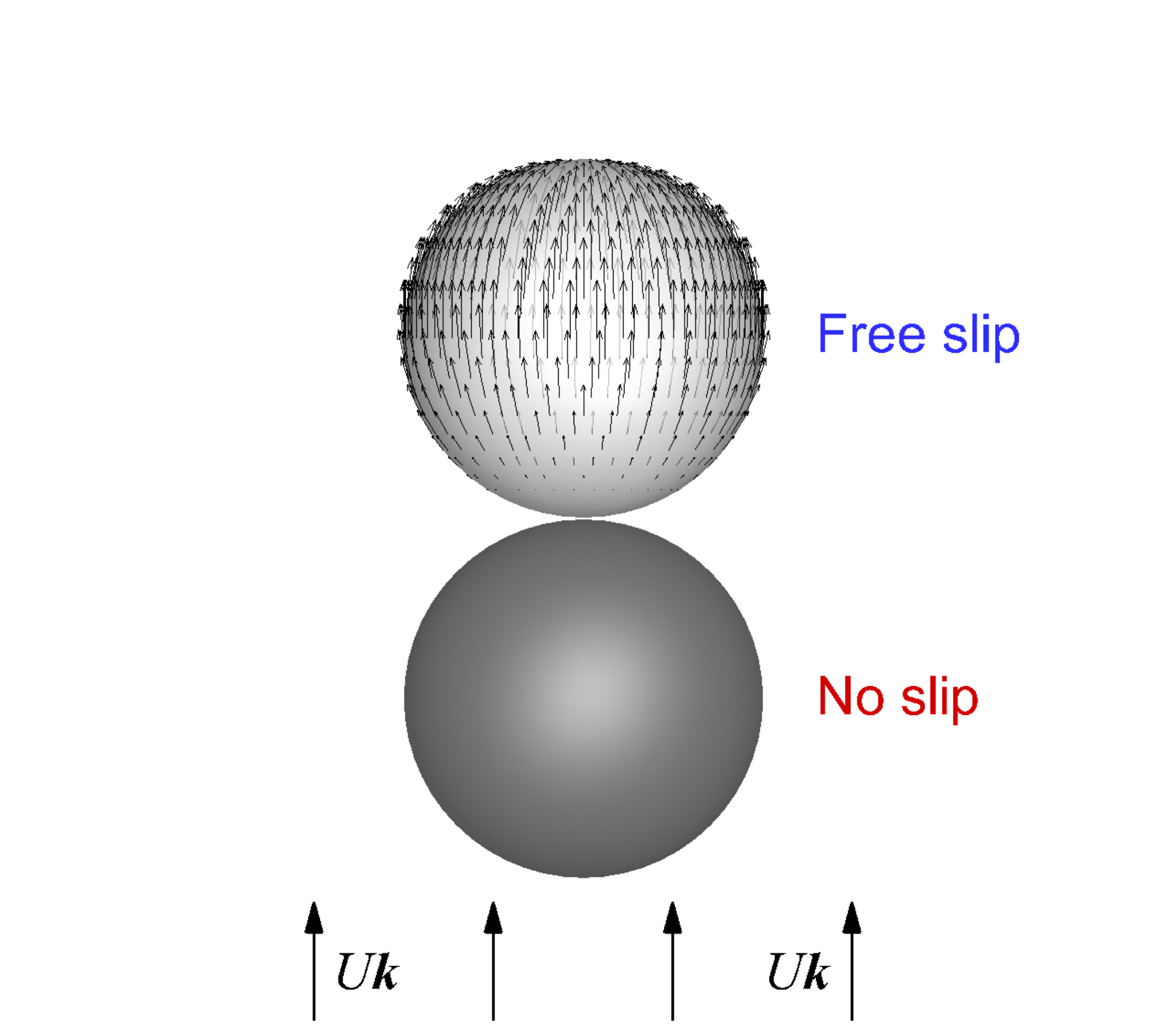} }
\caption{The surface velocity field, $\boldsymbol{u}$ on (a) two free slip spheres ($|u/U|_{max}=0.42$), (b) a free slip and a no slip sphere ($|u/U|_{max}=0.35$), at $h/R$ = 0.01 in a uniform external field, $U\boldsymbol{k}$ parallel to the line of centers. See Fig. 4 for details.}
\end{figure}}
}

{  % -- Figure 6

{\begin{figure}[t]
\centering
\subfloat[]{ \includegraphics[width=0.31\textwidth] {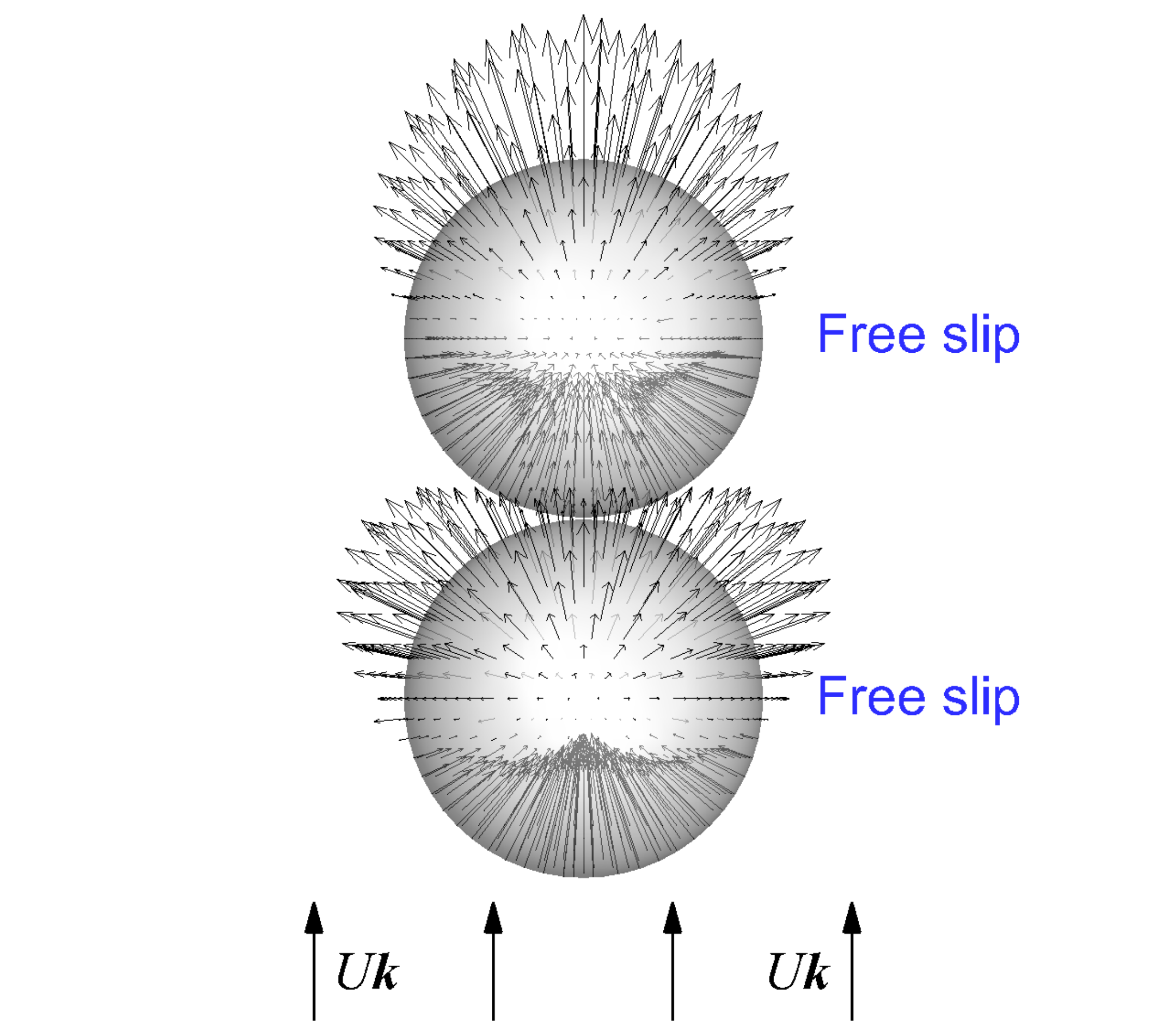} }
\subfloat[]{ \includegraphics[width=0.31\textwidth] {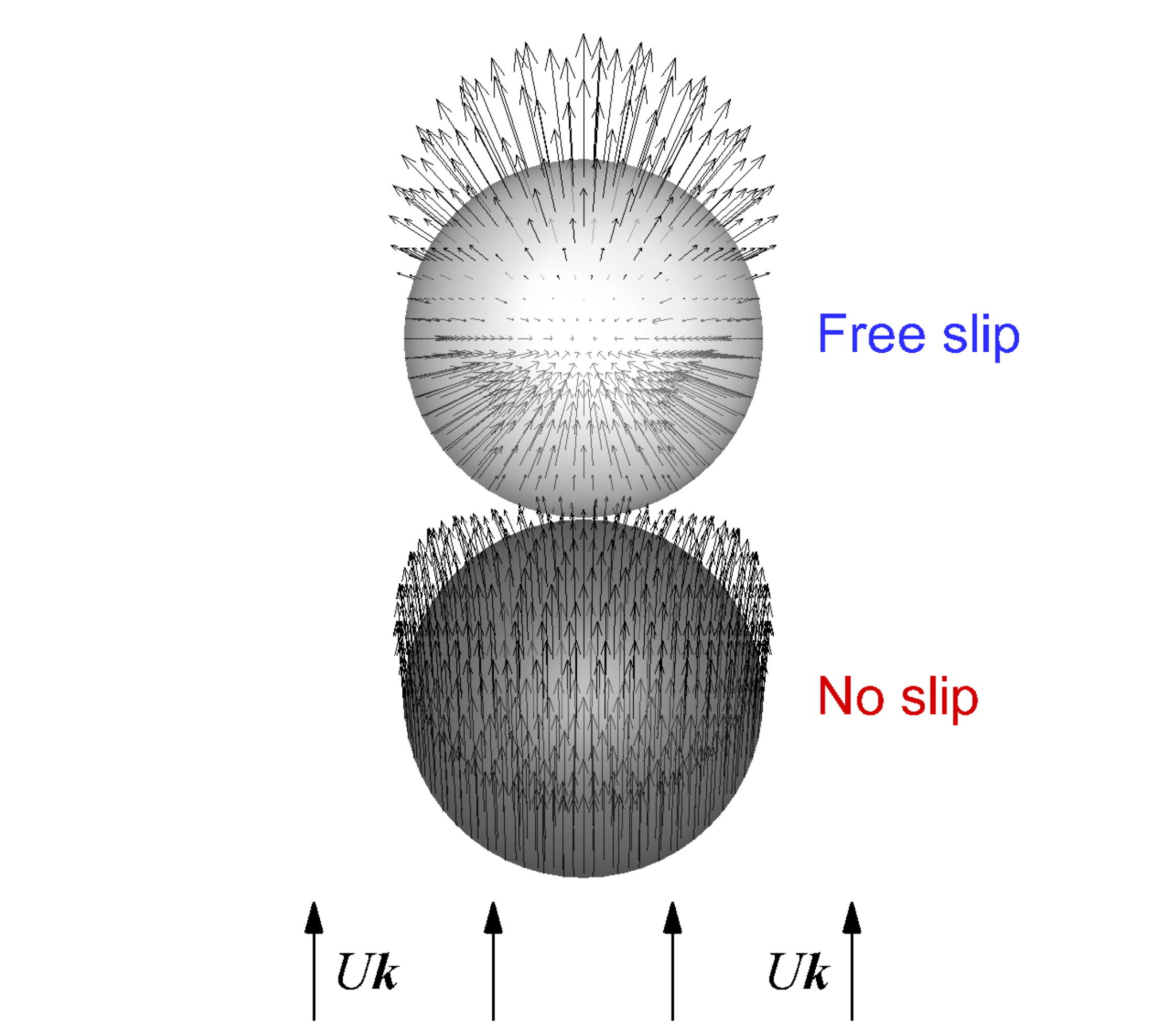} }
\subfloat[]{ \includegraphics[width=0.31\textwidth] {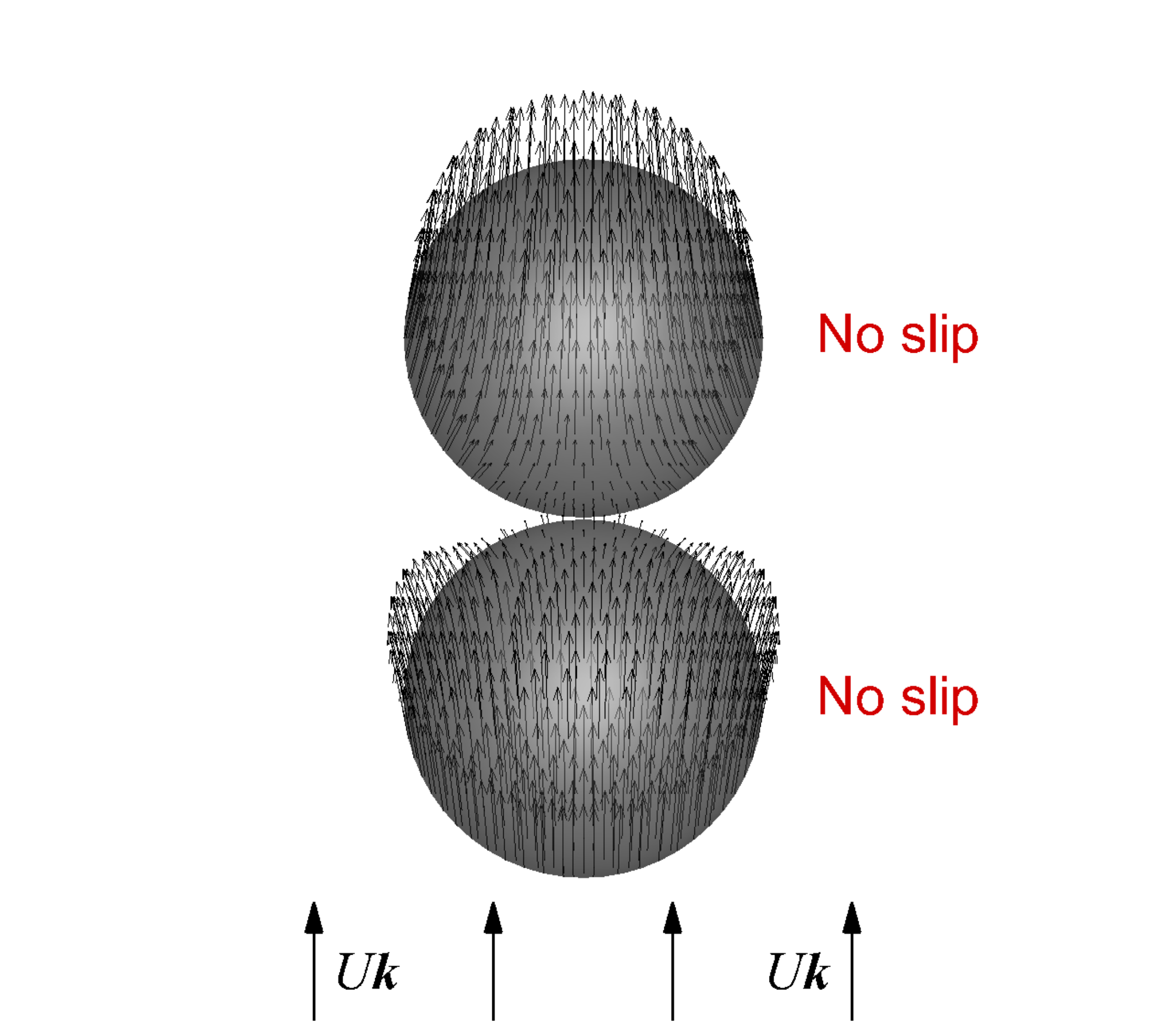} }
\caption{The traction field, $\boldsymbol{f}$ on (a) two free slip spheres ($|fR/\mu U|_{max}=2.53$), (b) a free slip and a no slip sphere ($|fR/\mu U|_{max}=2.19$), (c) two no slip spheres ($|fR/\mu U|_{max}=1.25$) at $h/R$ = 0.01 in a uniform external field, $U\boldsymbol{k}$ in the $z$-direction parallel to the line of centers. See Fig. 4 for details.}
\end{figure}}
}

In Figs. 7-9, we show similar results for the case when the uniform external flow field, $U\boldsymbol{i}$, is oriented perpendicular to the line of centers between the spheres. The drag force on each sphere in the direction of the external flow and the magnitude of the torque about the $y$-axis are shown in Fig. 7 as a function of sphere separation, $h/R$. Again the agreement with results from available exact analytic solutions\cite{ONeill1970} is excellent. The surface velocity vector field for the free slip/free slip and free slip/no slip combinations and the traction field for all three combinations at $h/R=0.01$ are shown in Figs. 8 and 9.

A further test of the numerical precision of our implementation is to compute components of the forces and torques in the above examples that are expected to be zero from symmetry considerations. We found that the magnitudes of these components are of the order $10^{-10}$ times smaller than the non-zero component. These results illustrate the flexibility and accuracy for all types of boundary conditions.

{  % -- Figure 7
{\begin{figure}[t]
\centering
\subfloat[]{ \includegraphics[width=0.47\textwidth] {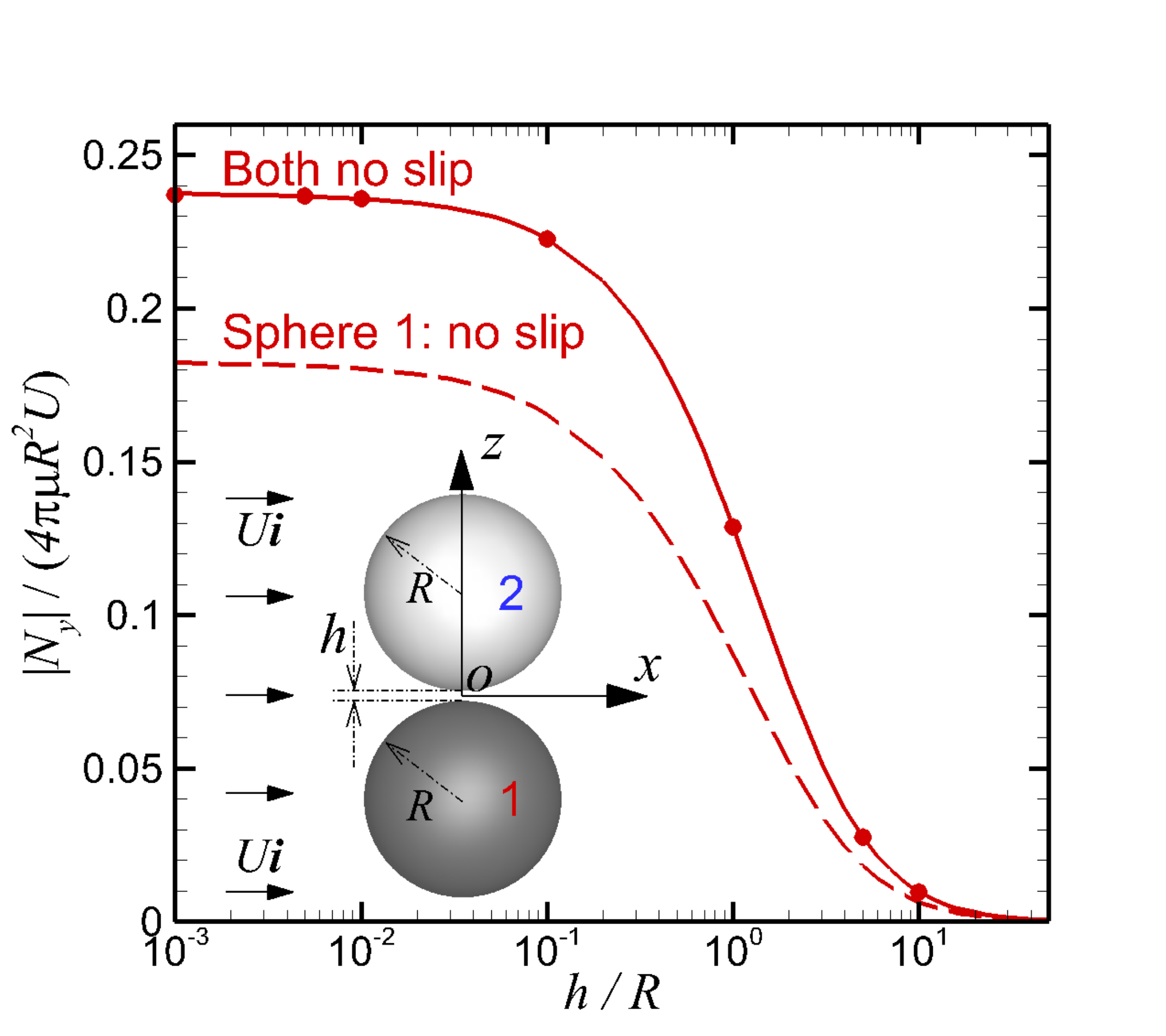} }
\subfloat[]{ \includegraphics[width=0.47\textwidth] {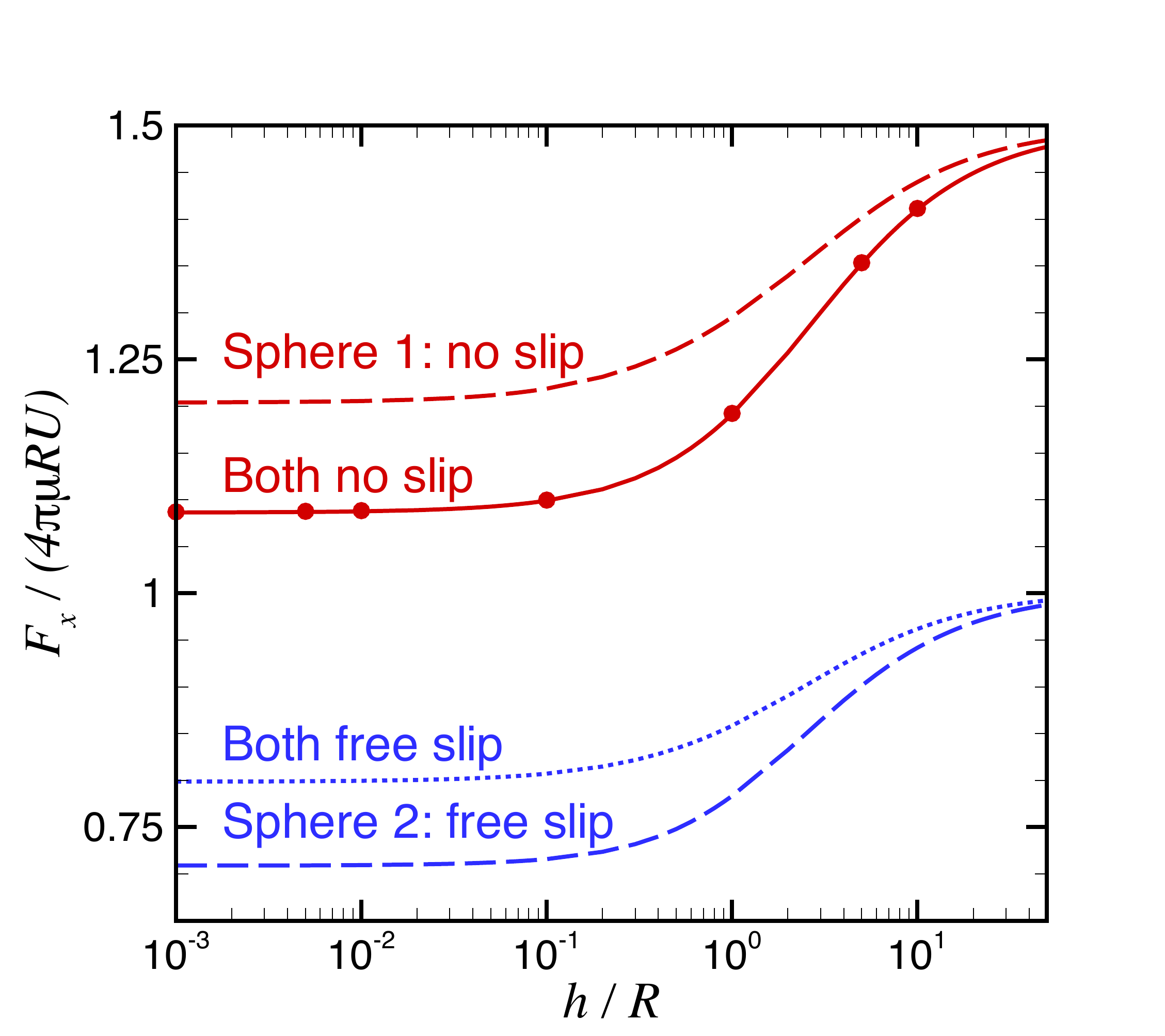} }
\caption{Torques and forces on two spheres in an external uniform flow field, $U\boldsymbol{i}$, in the $x$-direction perpendicular to the line of centers as functions of separation, $h/R$. (a) The magnitude of the torque about the $y$-axis, $N_y$, on both no slip spheres or on one no slip sphere. Inset: The geometry of two spheres in an uniform flow perpendicular to the line of centers, (b) The drag force in the $x$-direction, $F_x$ when one or both spheres are no slip or free slip, together with exact analytical solution for two no slip spheres ($\bullet$) \cite{ONeill1970}.  The BRIEF results (lines) with maximum relative error of less than 0.2\% are obtained using 2352 linear elements with 1178 nodes on each sphere and the source potential for $\boldsymbol{w}$.}
\end{figure}}
}

{  % -- Figure 8
{\begin{figure}[t]
\centering
\subfloat[]{ \includegraphics[width=0.31\textwidth] {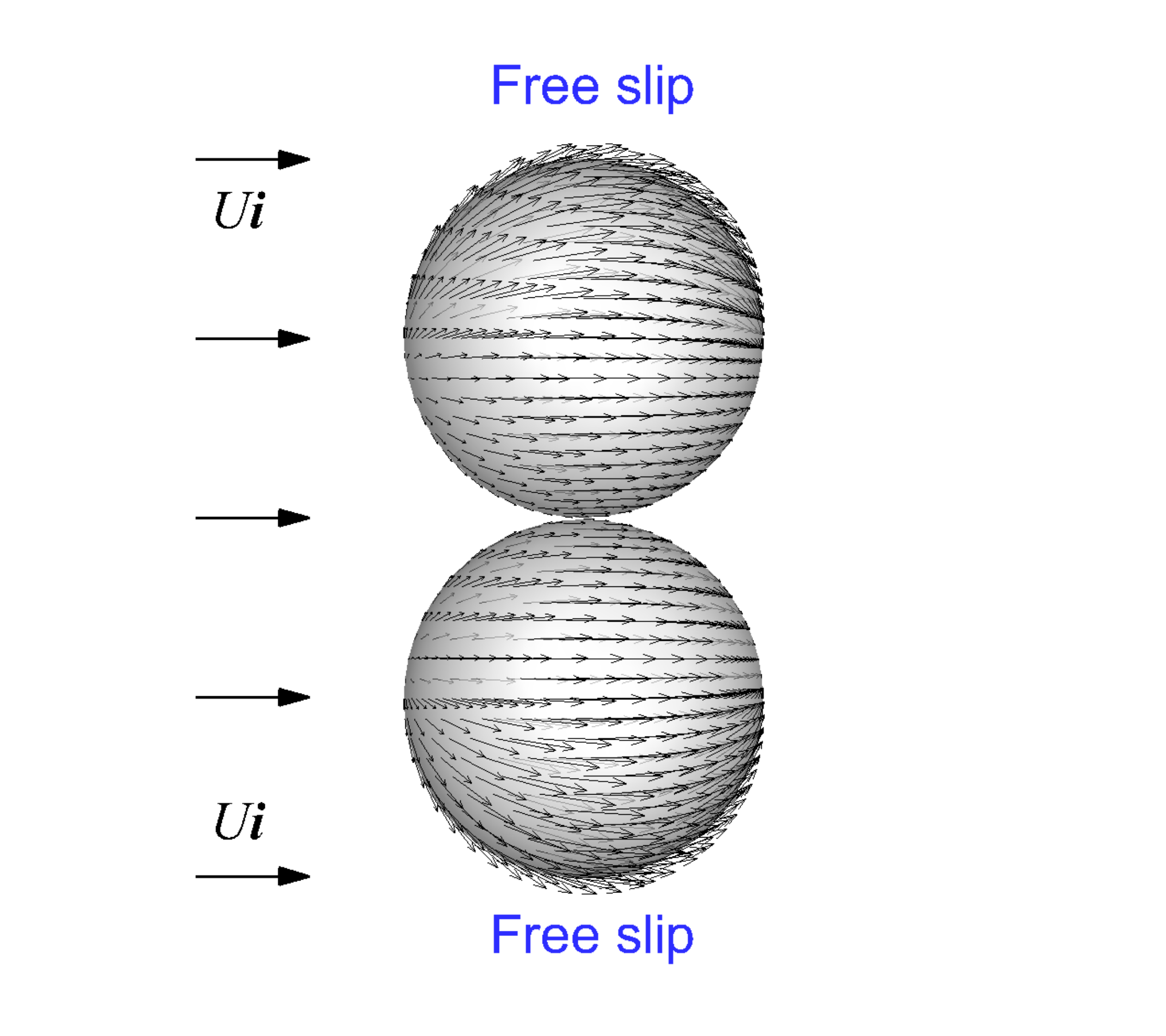} }
\subfloat[]{ \includegraphics[width=0.31\textwidth] {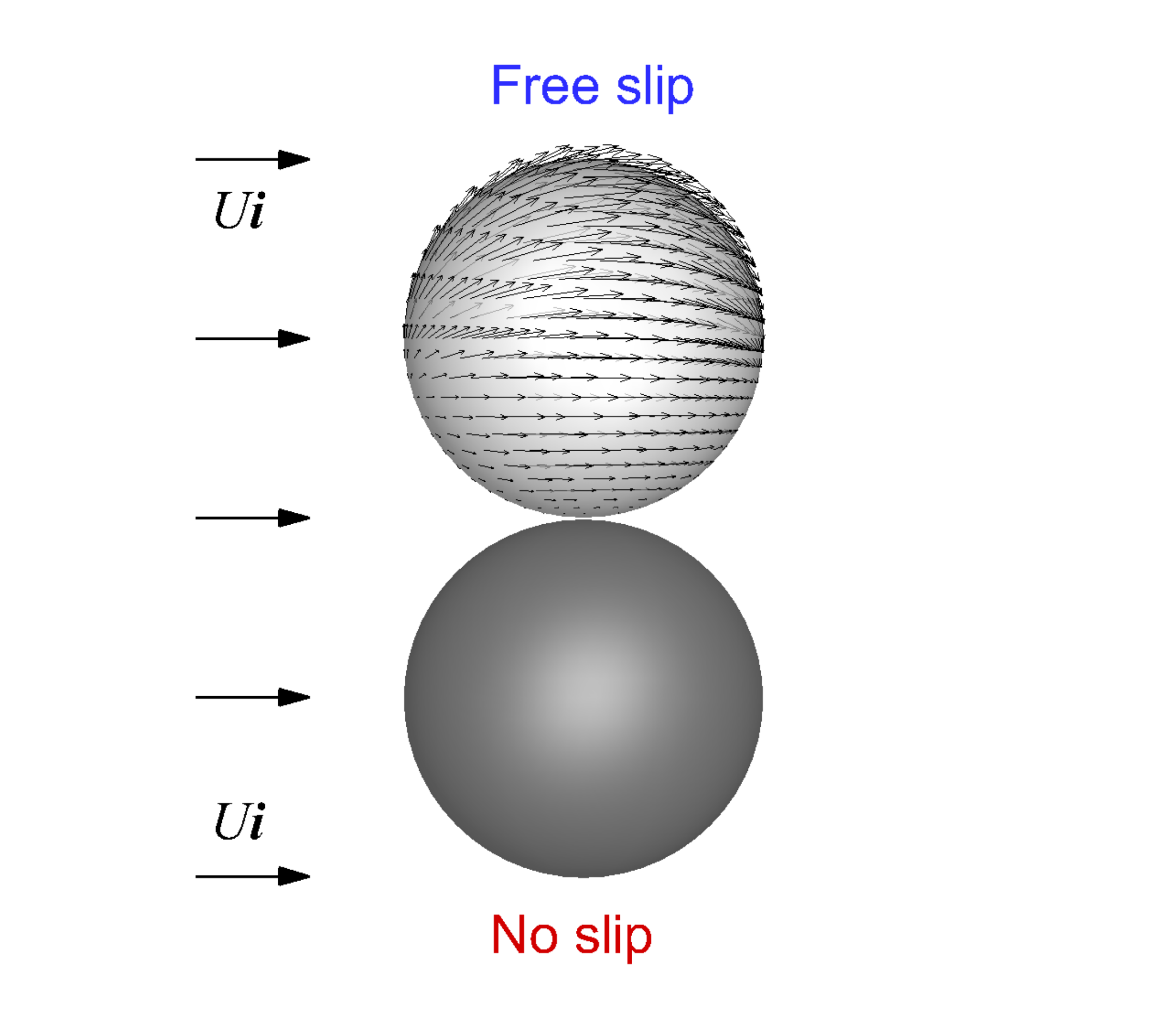} }
\caption{The surface velocity field, $\boldsymbol{u}$, on (a) two free slip spheres ($|u/U|_{max}=0.46$), (b) a free slip and a no slip sphere ($|u/U|_{max}=0.44$), at $h/R$ = 0.01 in a uniform external field, $U\boldsymbol{i}$, perpendicular to the line of centers. See Fig. 7 for details.}
\end{figure}}
}

{  % -- Figure 9
{\begin{figure}[t]
\centering
\subfloat[]{ \includegraphics[width=0.31\textwidth] {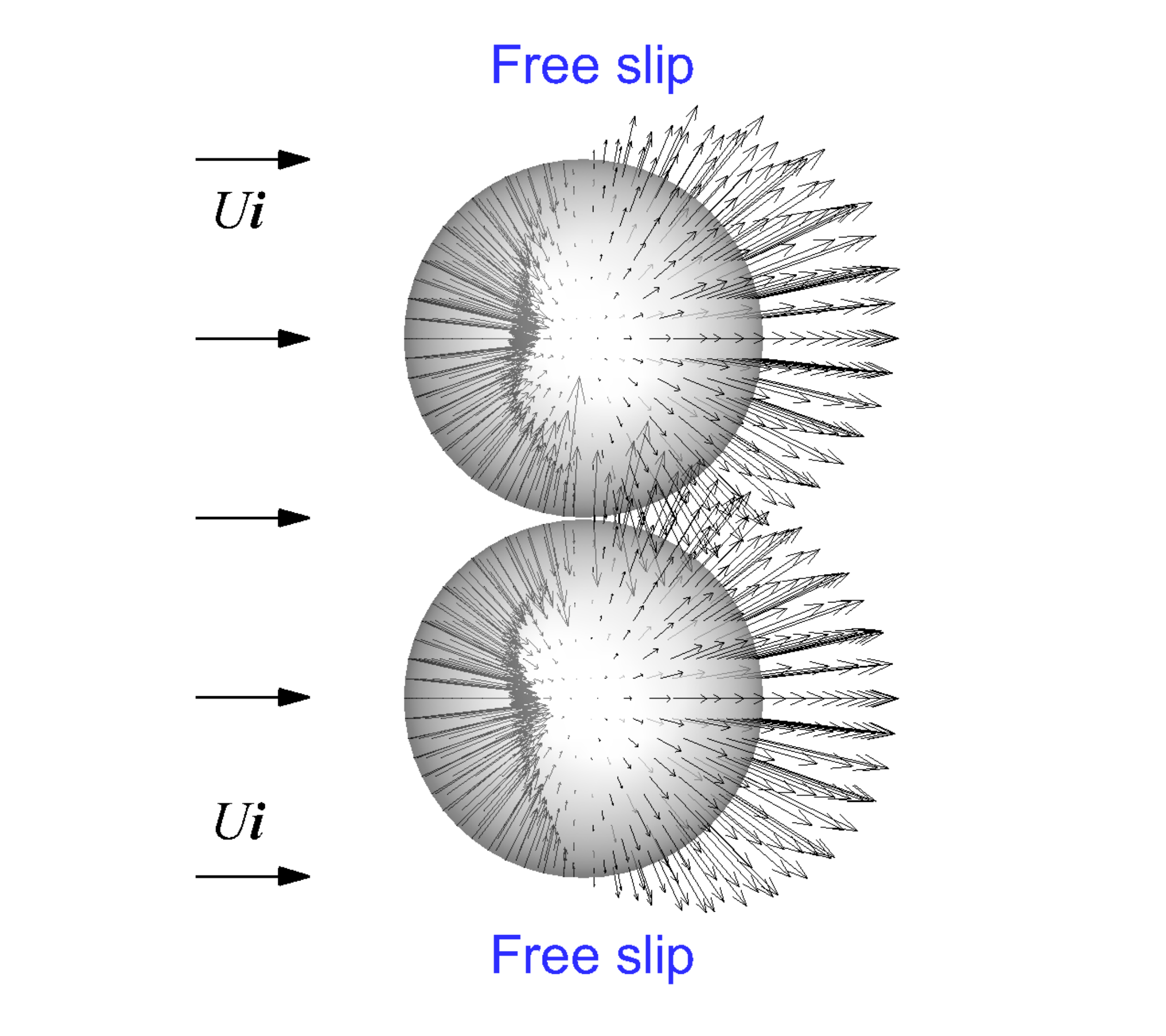} }
\subfloat[]{ \includegraphics[width=0.31\textwidth] {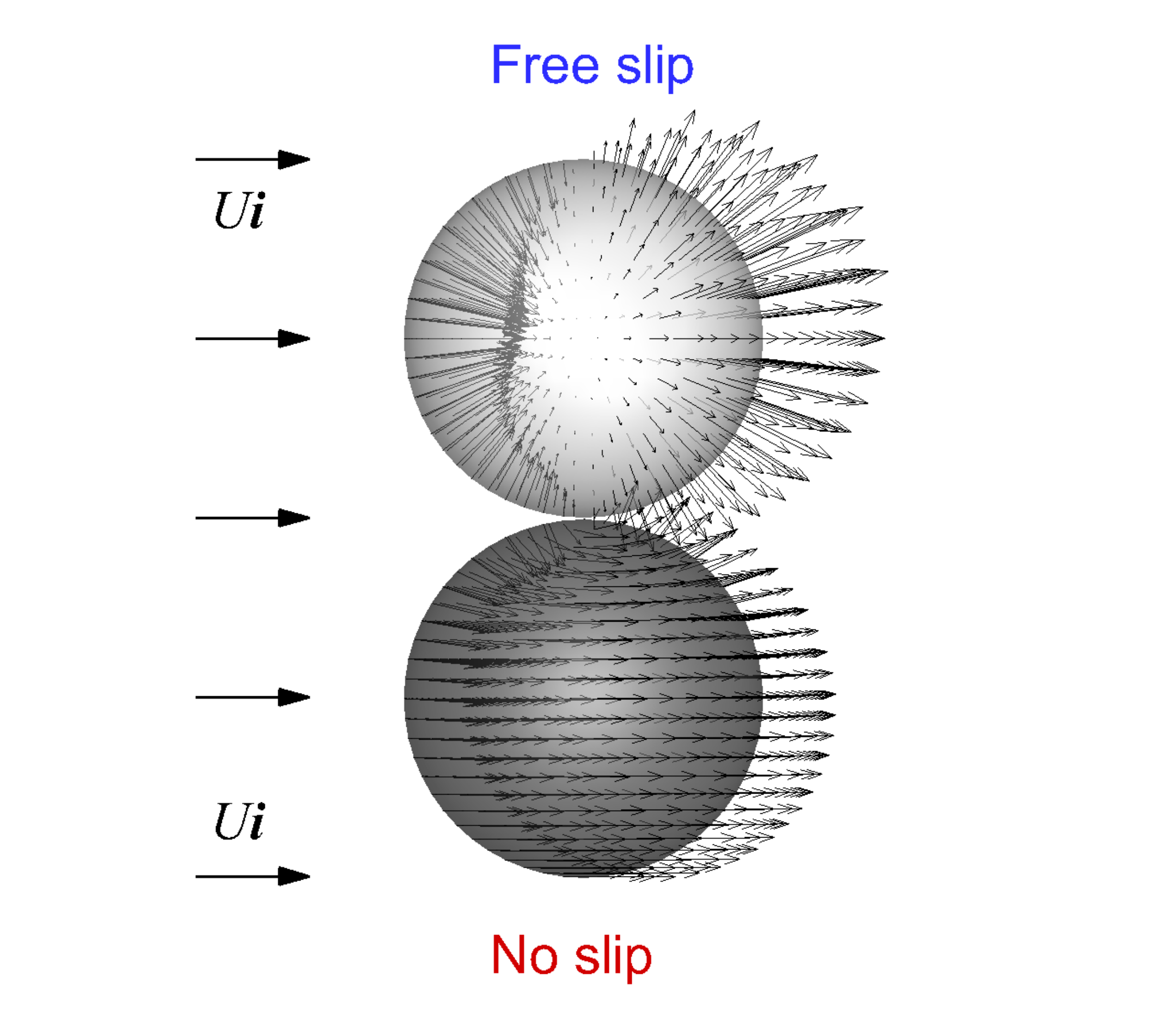} }
\subfloat[]{ \includegraphics[width=0.31\textwidth] {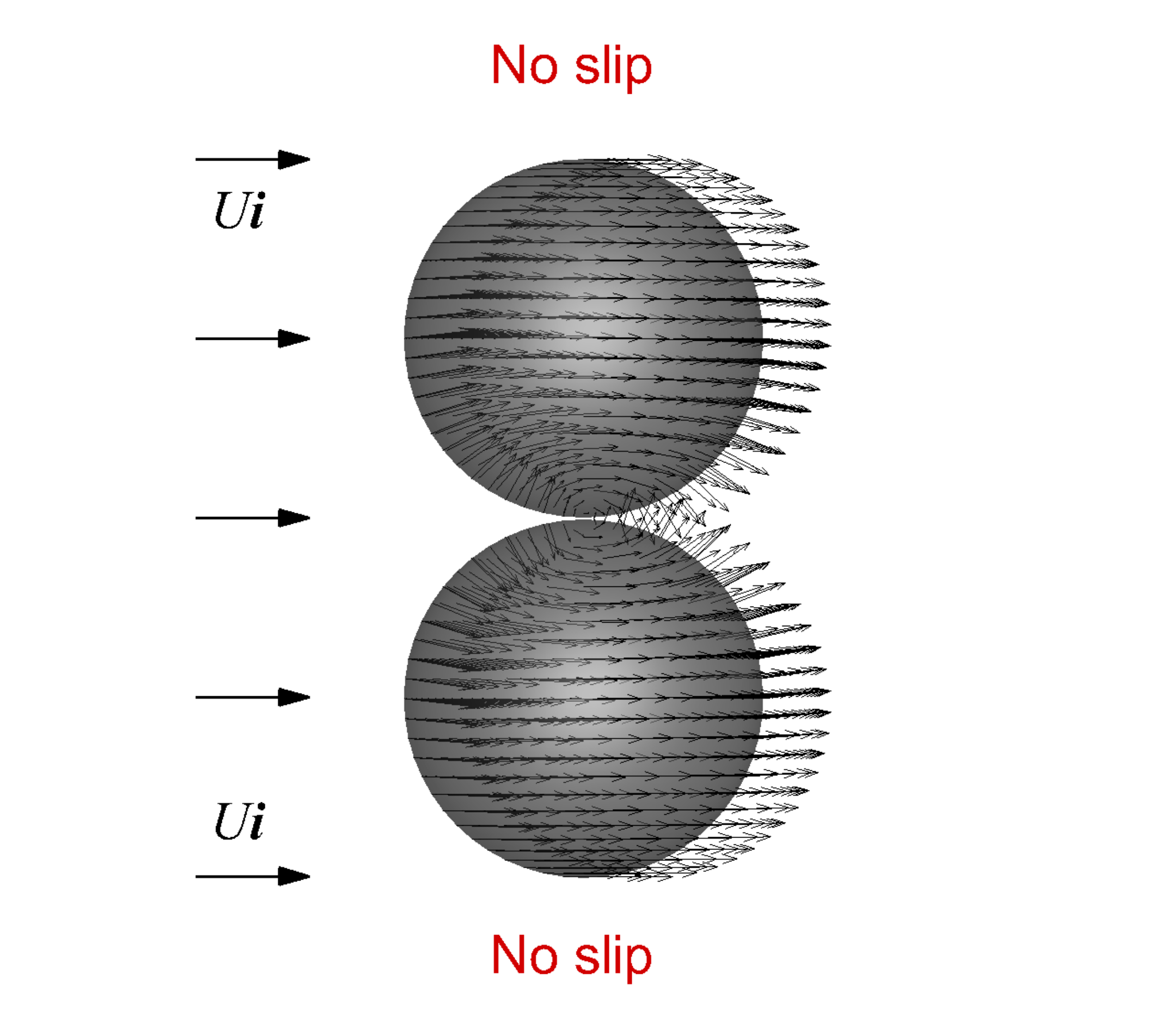} }
\caption{The traction field, $\boldsymbol{f}$, on (a) two free slip spheres ($|fR/\mu U|_{max}=2.51$), (b) a free slip and a no slip sphere ($|fR/\mu U|_{max}=2.30$), (c) two no slip spheres ($|fR/\mu U|_{max}=1.37$) at $h/R$ = 0.01 in a uniform external field, $U\boldsymbol{i}$, in the $x$-direction perpendicular to the line of centres. See Fig. 7 for details.}
\end{figure}}
}

% -- Figure 10
{\begin{figure}[t]
\centering
\subfloat[]{ \includegraphics[width=0.47\textwidth] {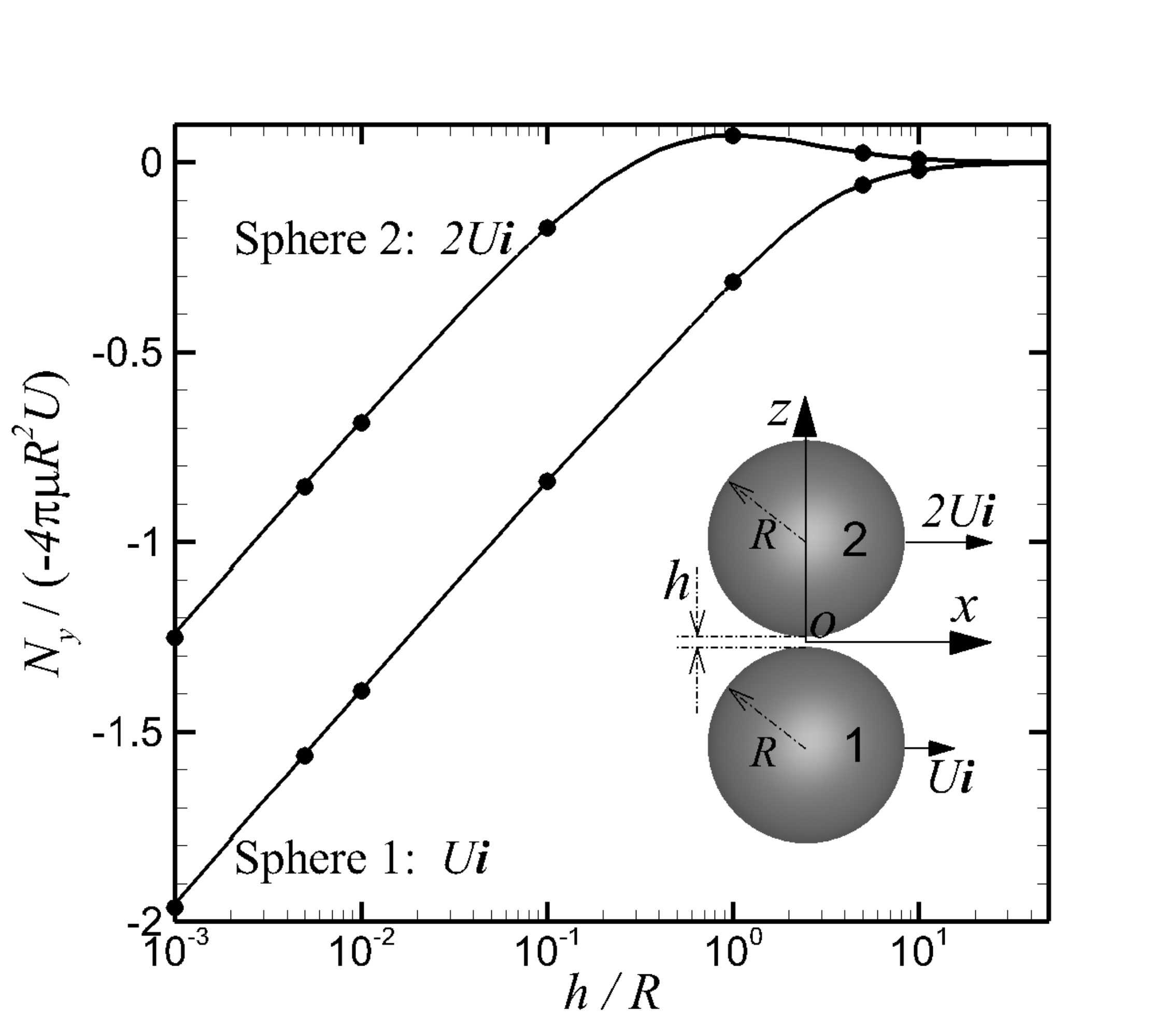} }
\subfloat[]{ \includegraphics[width=0.47\textwidth] {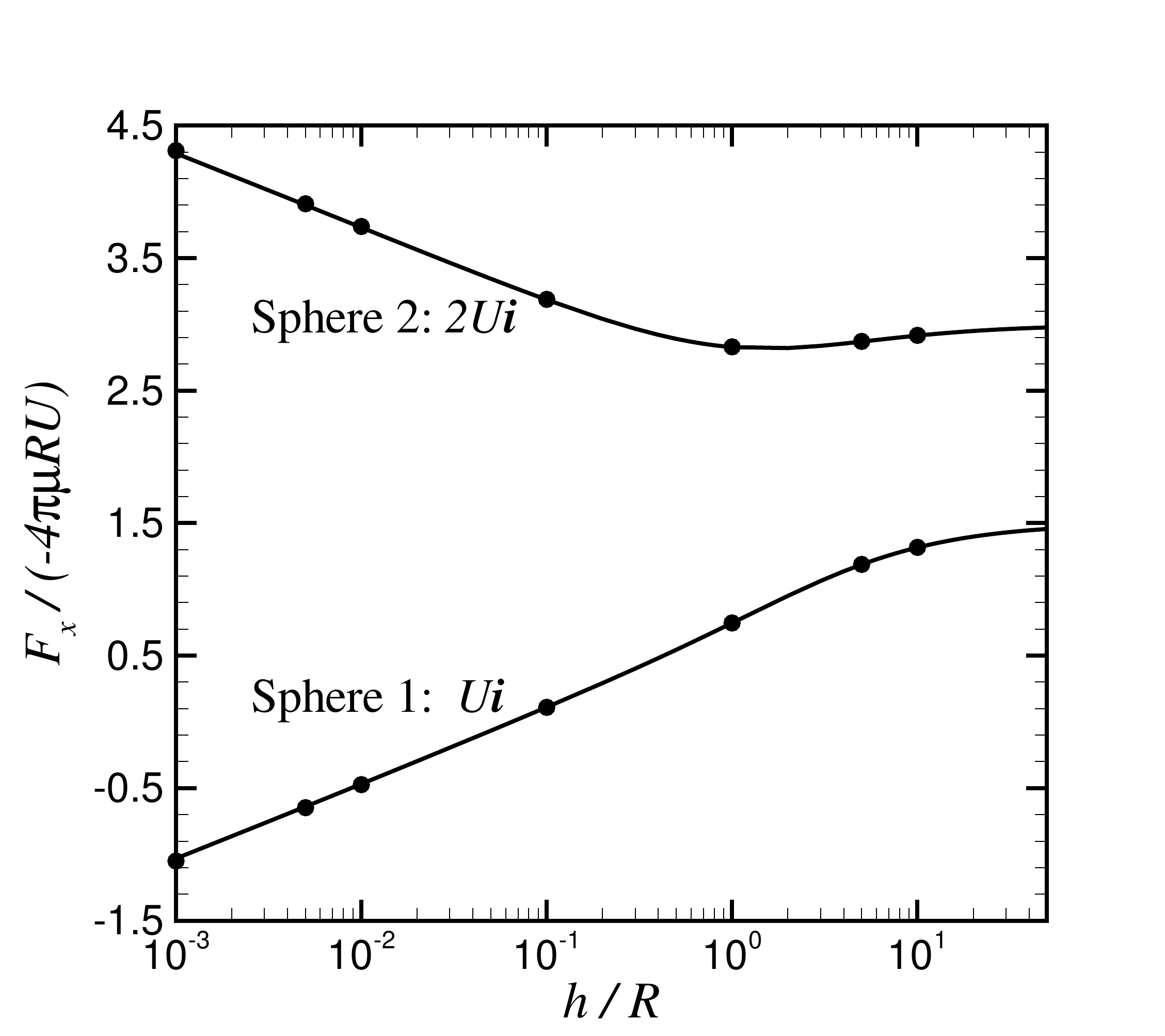} }
\caption{Torques and forces on two no-slip spheres in a stationary flow field moving with different velocity in the $x$-direction perpendicular to the line of centers as functions of separation, $h/R$. (a) The magnitude of the torques about the $y$-axis, $N_y$. Inset: The geometry of two spheres, (b) The drag force in the $x$-direction, $F_x$, compared with the exact bipolar asymptotic solution $(\bullet)$\cite{ONeill1970}. The BRIEF results with a maximum relative error less than 1.8\% are obtained using 4800 linear triangular elements connected by 2402 nodes on each sphere with the source potential for $\boldsymbol{w}$.}
\end{figure}}

\subsection{Lubrication flow}
Problems in which lubrication effects within thin gaps between closely spaced boundaries are important provide additional challenges to the boundary integral methods. In addition to the mathematical singularities in the conventional boundary integral method, the physics of the problem can give rise to divergences such as an unbounded pressure field as the gap width approaches zero. Although the BRIEF removed the mathematical singularities, the divergent behavior of the relevant physical quantities must remain.

% -- Figure 11
{\begin{figure}[t]
\centering
\subfloat[]{ \includegraphics[width=0.47\textwidth] {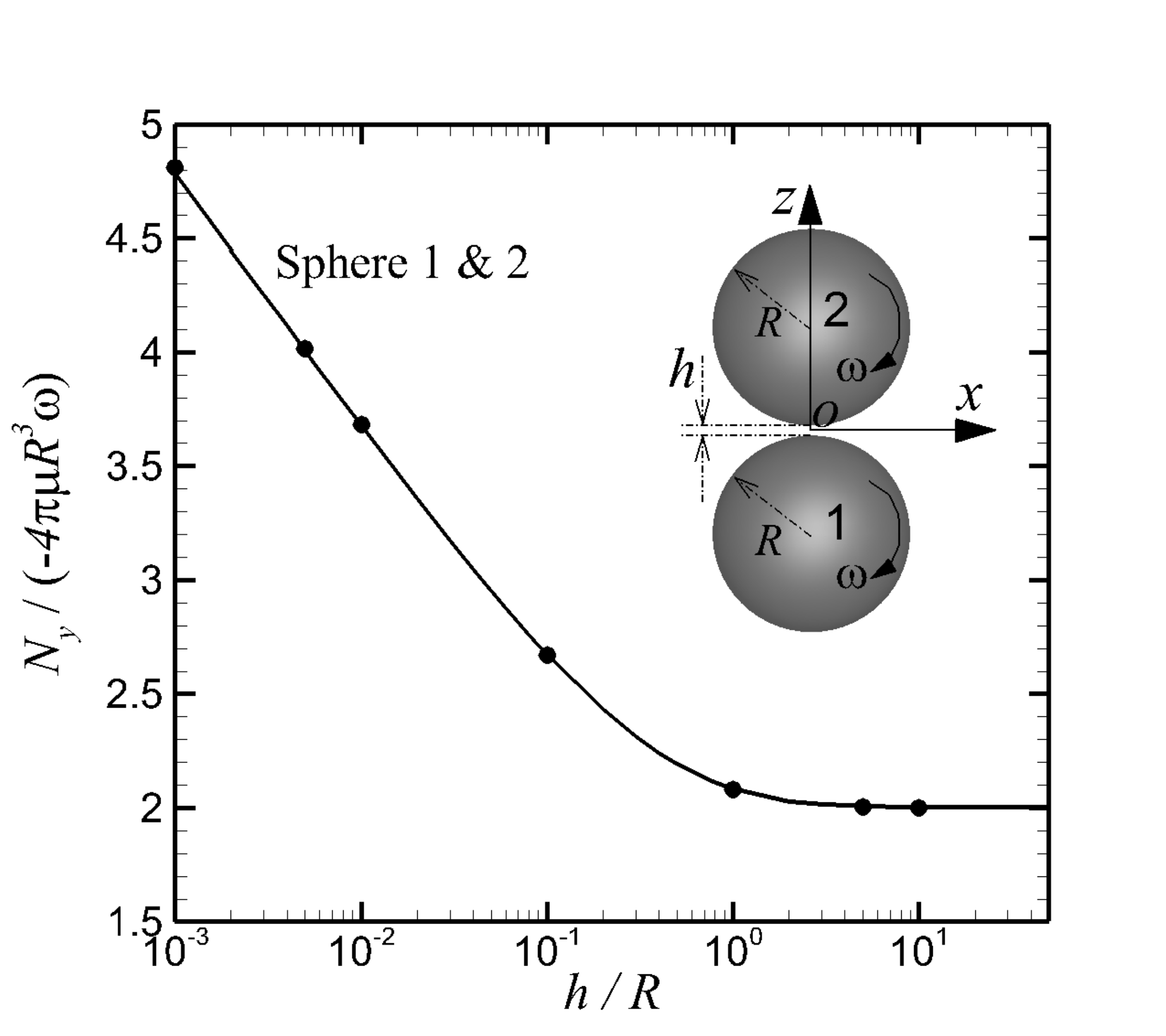} }
\subfloat[]{ \includegraphics[width=0.47\textwidth] {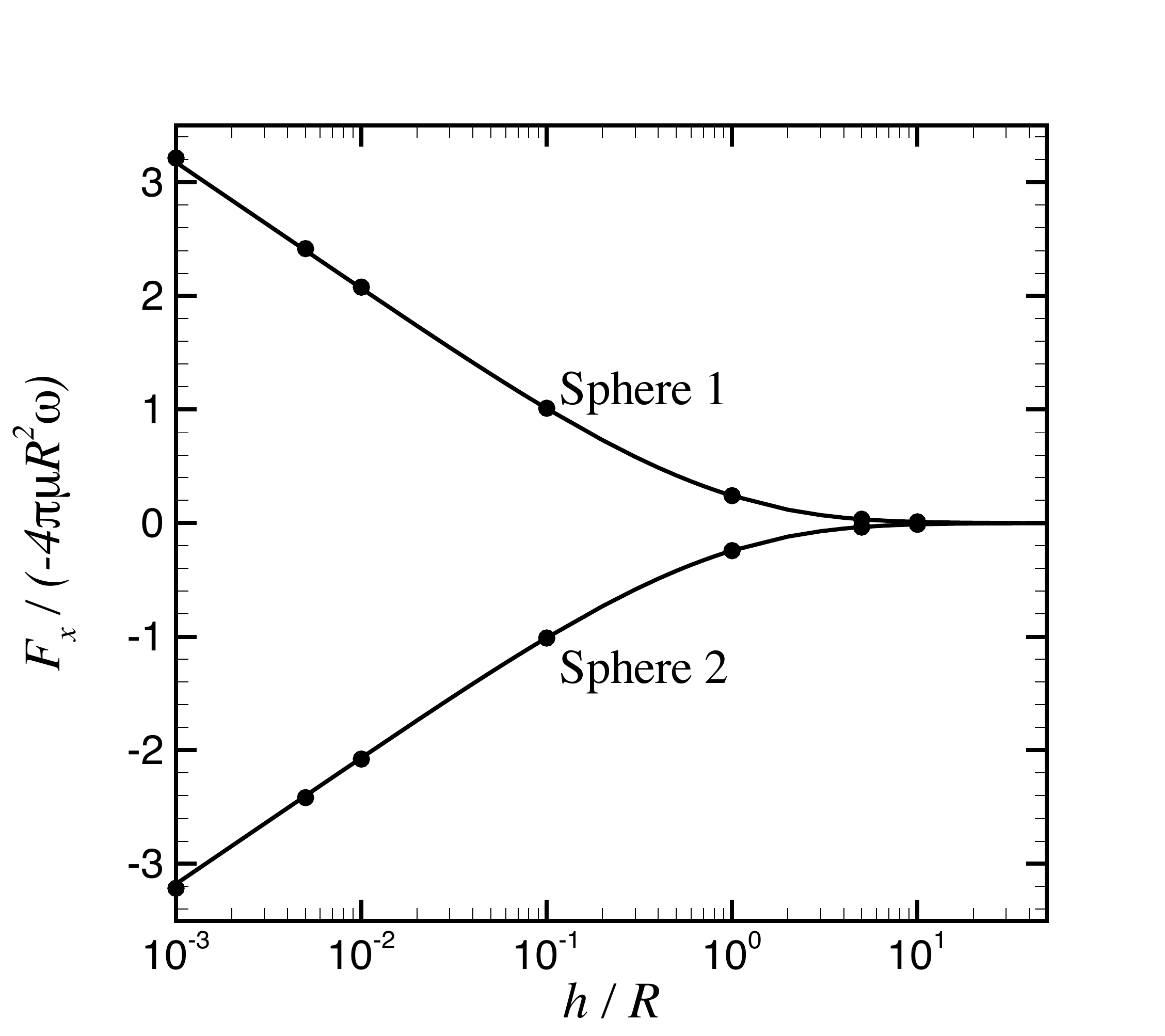} }
\caption{Torques and forces on two spheres no-slip spheres in a stationary flow field rotating in the $y$-direction perpendicular to the line of centers as functions of separation, $h/R$. (a) The magnitude of the torque about the $y$-axis, $N_y$. Inset: The geometry of two spheres, (b) The drag force in the $x$-direction, $F_x$, compared with the exact analytical solution $(\bullet)$\cite{ONeill1970}. The BRIEF results with a maximum relative error less than 1.2\% are obtained using 4800 linear triangular elements connected by 2402 nodes on each sphere with the source potential for $\boldsymbol{w}$.}
\end{figure}}

To illustrate such situations, we consider two examples: two nearly touching spheres with no-slip boundary condition moving with different velocities, $U$ and $2U$, perpendicular to their line of centers (Fig. 10a), and two nearly touching spheres rotating with the same angular velocity, $\omega$, about axes perpendicular to their line of centers (Fig. 11a). Due to the lubrication effect, the pressure within the gap becomes very high as the sphere separation decreases, and can become challenging numerically. To obtain satisfactory results corresponding to a minimum sphere separation, scaled by the radius, $h/R$ down to $10^{-3}$,  we employ a finer uniform mesh with 4800 linear triangular elements connected by 2402 nodes on each sphere - about twice as dense as previous examples. The results obtained using BRIEF with the source potential (Item IV in Table I) are in excellent agreement with exact bipolar coordinate system solution,\cite{ONeill1970} as shown in Figs. 10 and 11.

\subsection{Forces and torques on spheroids}

As a final example, we consider forces experienced by a solid prolate spheroid translating with constant velocity, $U$, along its axes as functions of the aspect ratio and varying Navier slip boundary conditions (Fig. 12). We also investigate the torque experienced by a prolate spheroid rotating about the major axes at constant angular velocity, $\omega$ (Fig. 13). The surface of the spheroid, with semi-major and semi-minor axes $a$ and $b$, is defined by
\begin{align}\label{eq:prolatesph}
\frac{x^2}{a^2} + \frac{y^2+z^2}{b^2} = 1,\quad{b\le a}.
\end{align}
and is represented by a linear triangular mesh with 2352 elements connected by 1178 nodes. Analytical expressions for the forces and torques of such particles with the no slip boundary condition are given by Chwang and Wu.\cite{Chwang1974, Chwang1975} For comparison we also give results for spheroids with the Navier slip boundary condition for different slip lengths, $s$, in which, the analytical solutions are given by Lamb\cite{Lamb1932} when $a=b$ (sphere).

Results for the forces on the prolate spheroid in external flow fields of different orientations are given in Fig. 12 and for the torques on rotating prolate spheroids are given in Fig. 13. In those cases for which analytical results are available, the agreement with the numerical results obtained using BRIEF is excellent.

% -- Figure 12
{\begin{figure}[!hpt]
\centering
\subfloat[]{ \includegraphics[width=0.45\textwidth] {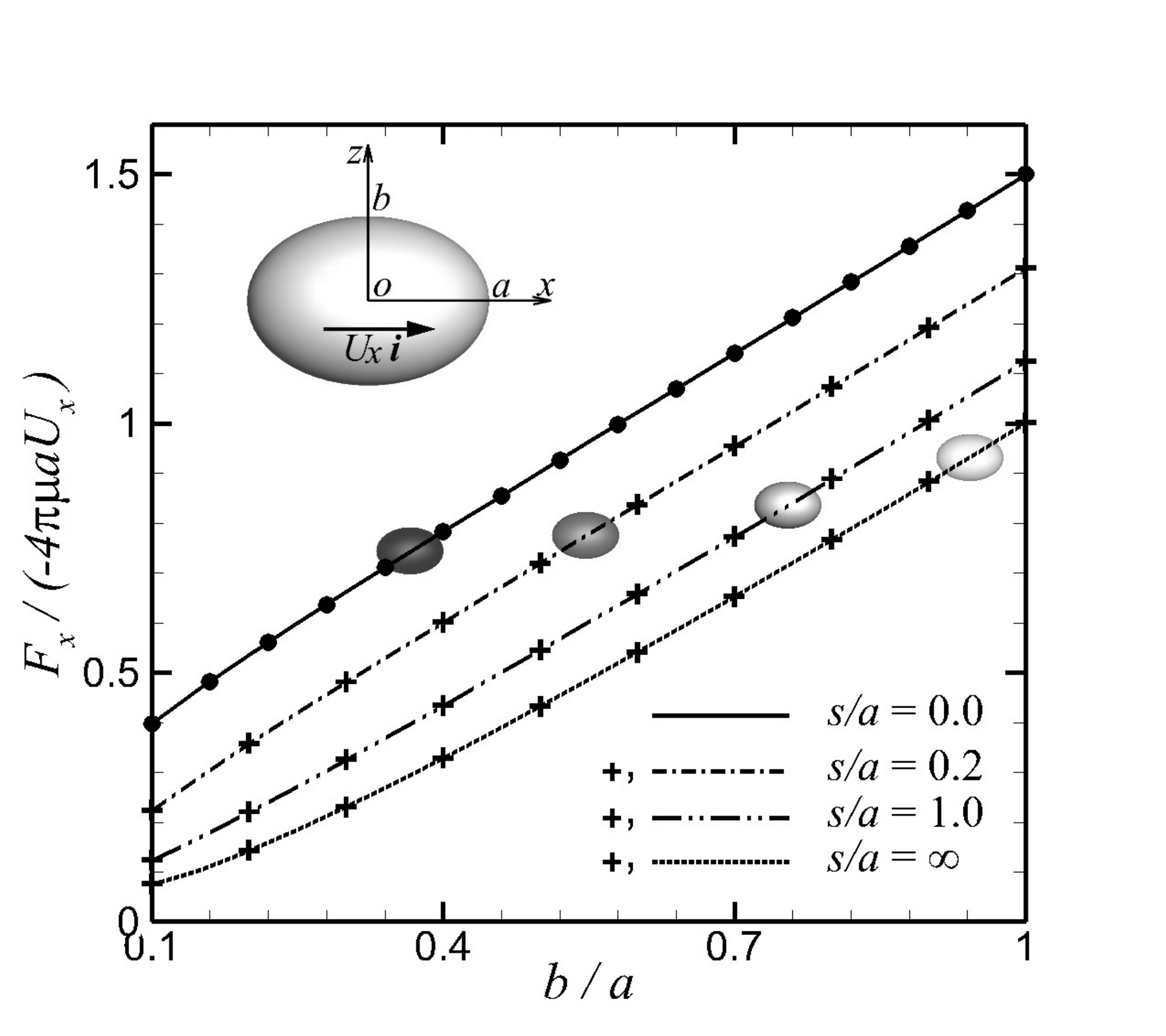} }
\subfloat[]{ \includegraphics[width=0.45\textwidth] {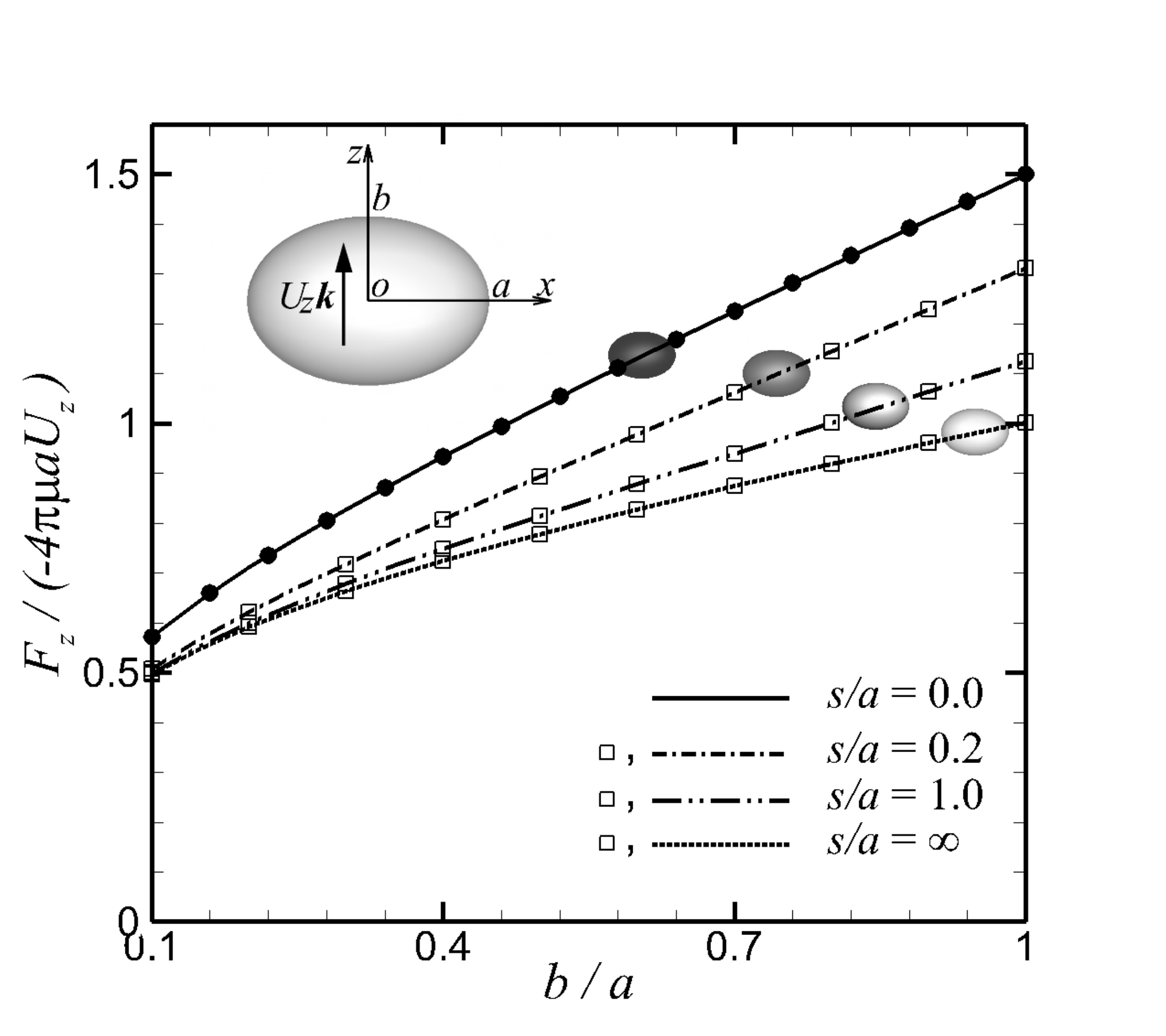} } \\
\subfloat[]{ \includegraphics[width=0.45\textwidth] {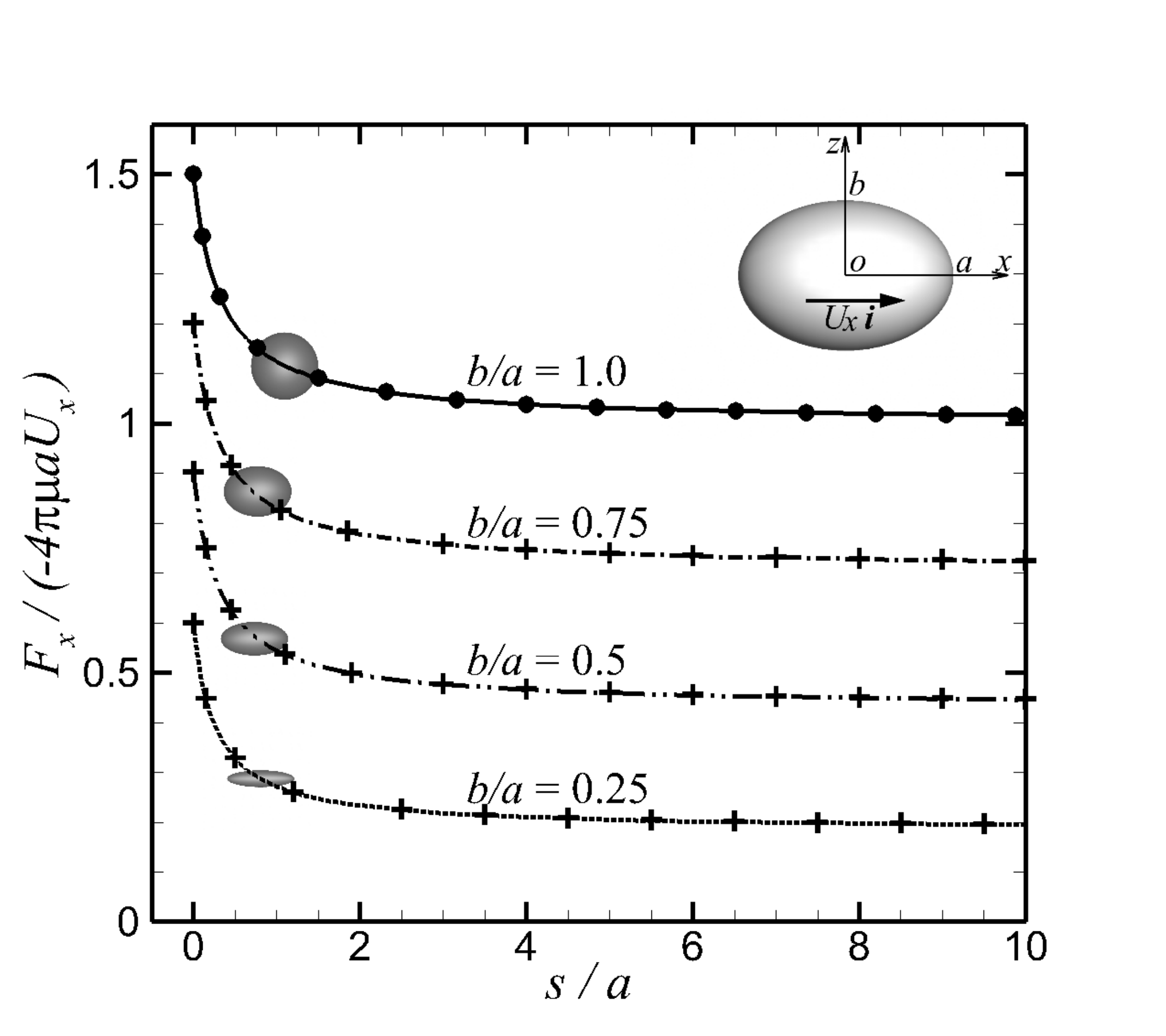} }
\subfloat[]{ \includegraphics[width=0.45\textwidth] {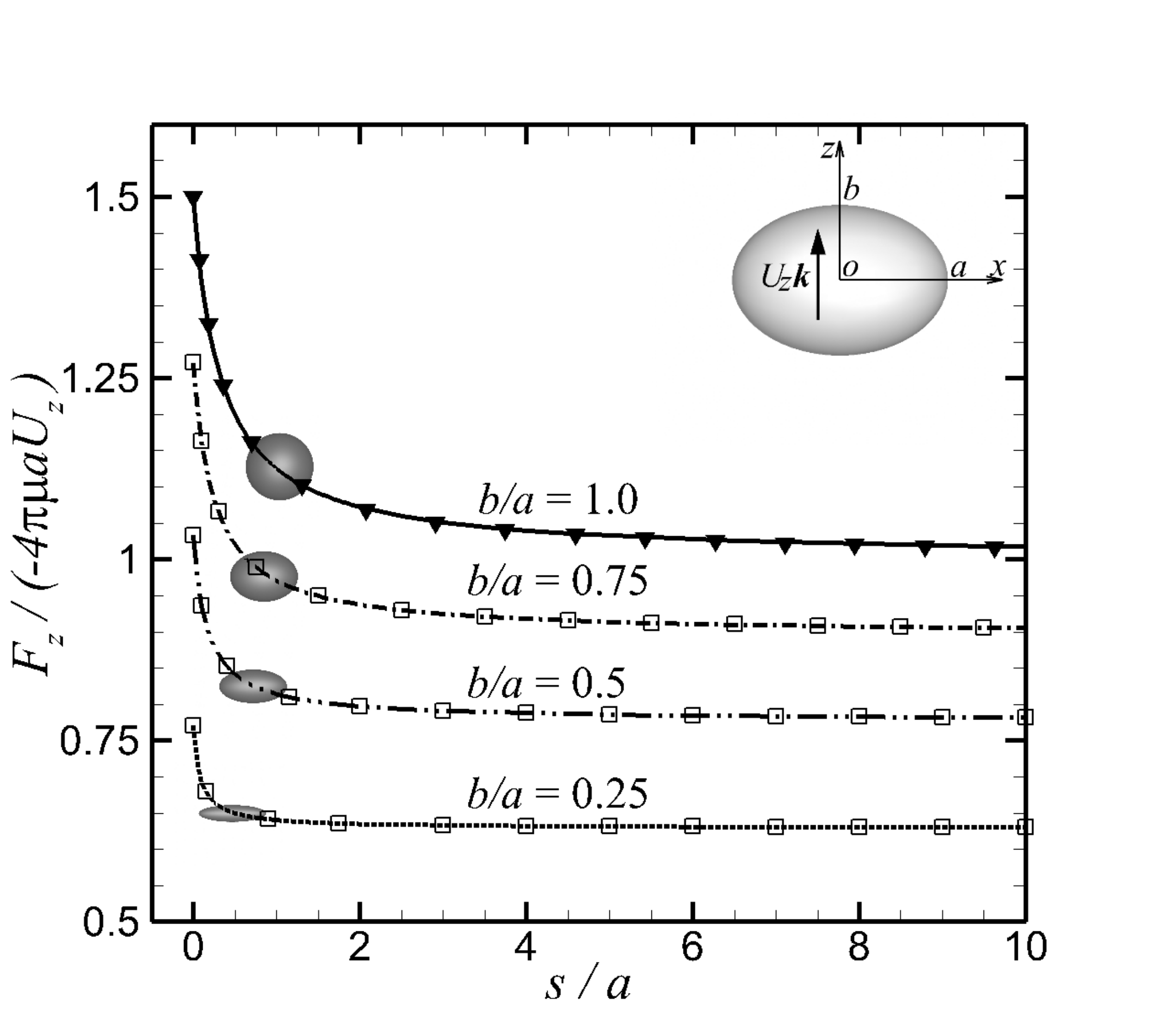} }
\caption{The drag force coefficients in (a) $x$-direction and (b) $z$-direction of the moving prolate spheroid as a function of $b/a$ in stationary Stokes flow; the drag force coefficients in (c) $x$-direction and (d) $z$-direction of the moving prolate spheroid as a function of the Navier slip length $s/a$ in stationary Stokes flow. Results from BRIEF with the linear function for $\boldsymbol{w}$ (Item I in Table I) are shown as lines. For comparison, results shown as $(+)$ in (a) and (c) are obtained by BRIEF with the source potential $\boldsymbol{w}$ (Item IV in Table I), and results shown as $(\Box)$ in (b) and (d) are obtained by BRIEF with the stresslet $\boldsymbol{w}$ (Item V in Table I). Analytical results are shown as $(\bullet)$\cite{Chwang1975} in (a) and (b) and $(\blacktriangledown)$\cite{Lamb1932} in (c) and (d). The BRIEF results with maximum relative error of less than 0.2\% are obtained using 2352 linear elements with 1178 nodes on the spheroid.}
\end{figure}}

%==================================================
% -- SEC 7 --
\section{Conclusions}

In this paper, we have shown how the Stokes equation for single-phase flow with prescribed boundary conditions can be cast as a boundary regularized integral equation (\ref{eq:stkflnsbie}) with integrands that are completely free of singularities that normally exist in the traditional form of the boundary integral equation. The singularities are removed by subtracting an auxiliary known flow field $\boldsymbol{w}$ that has the same singular behaviour as in the traditional formulation of the problem. As a result, the singularities as well as the solid angle that appear in the conventional boundary integral formulation are removed analytically without the need to introduce additional regularization parameters.

The present formulation is therefore fundamentally different to the regularized Stokeslet approach \cite{Cortez2001} that removes the singularities in the fundamental solutions by spreading a $\delta$-function force over a small ball of radius $\epsilon$, where the optimal value of $\epsilon$ that minimises error needs to be determined. The numerical effort in solving the linear system in BRIEF is comparable to that in CBIM but no special provisions are needed to compute the matrix coefficients.  In this sense, it is more efficient that the contour integral formulation~\cite{Bazhlekov2004}.

% -- Figure 13
{\begin{figure}[!hpt]
\centering
\subfloat[]{ \includegraphics[width=0.45\textwidth] {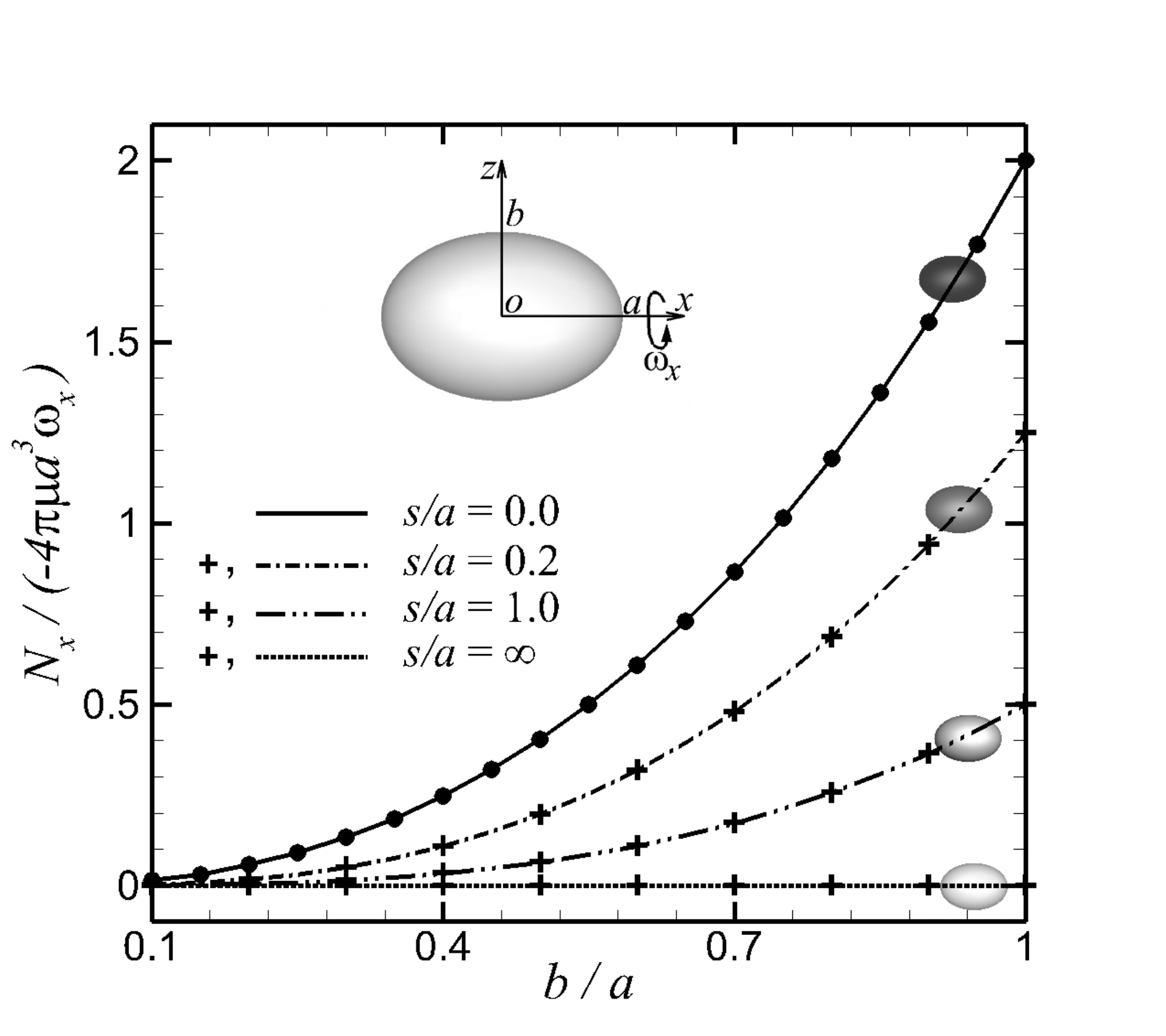} }
\subfloat[]{ \includegraphics[width=0.45\textwidth] {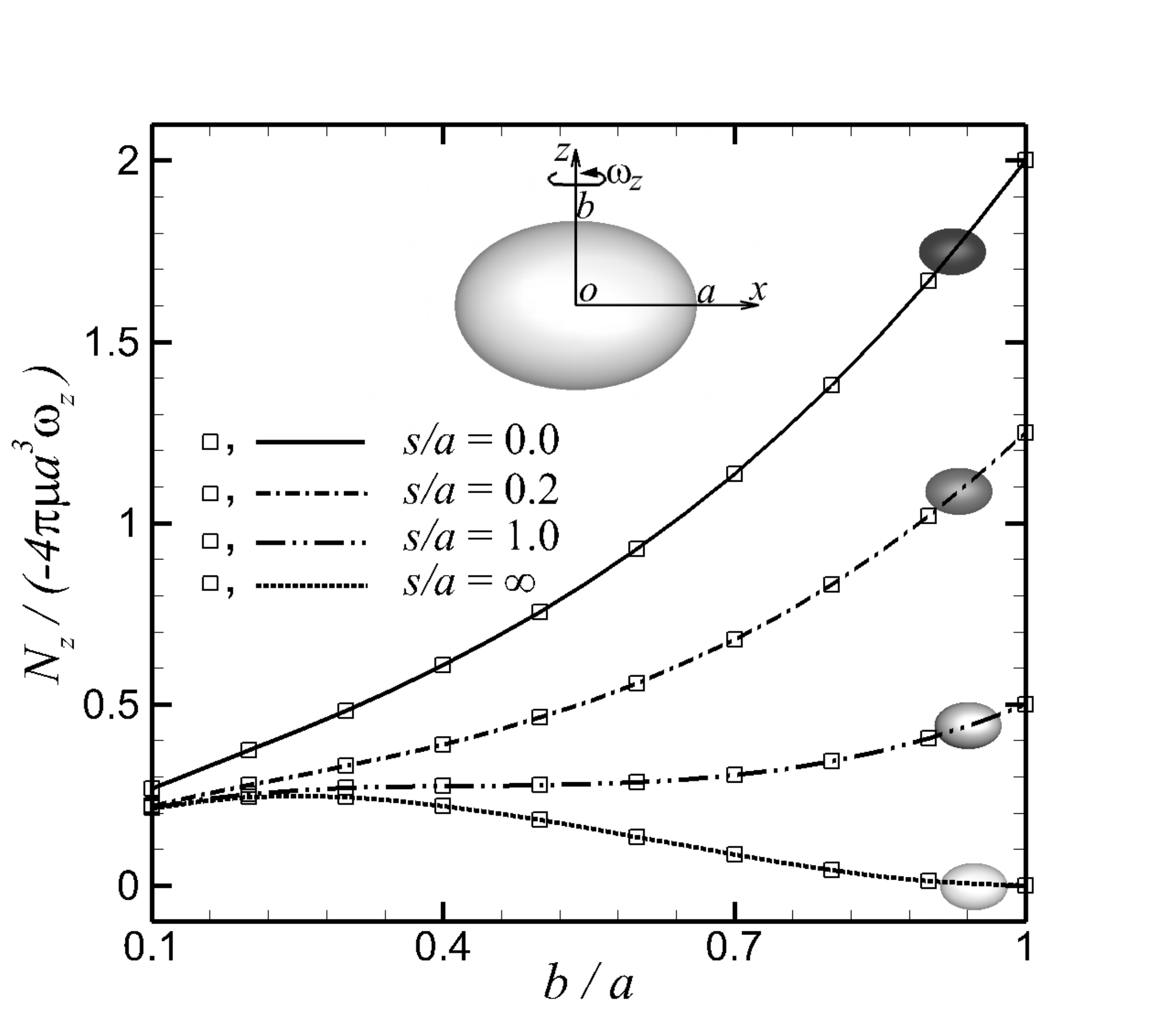} } \\
\subfloat[]{ \includegraphics[width=0.45\textwidth] {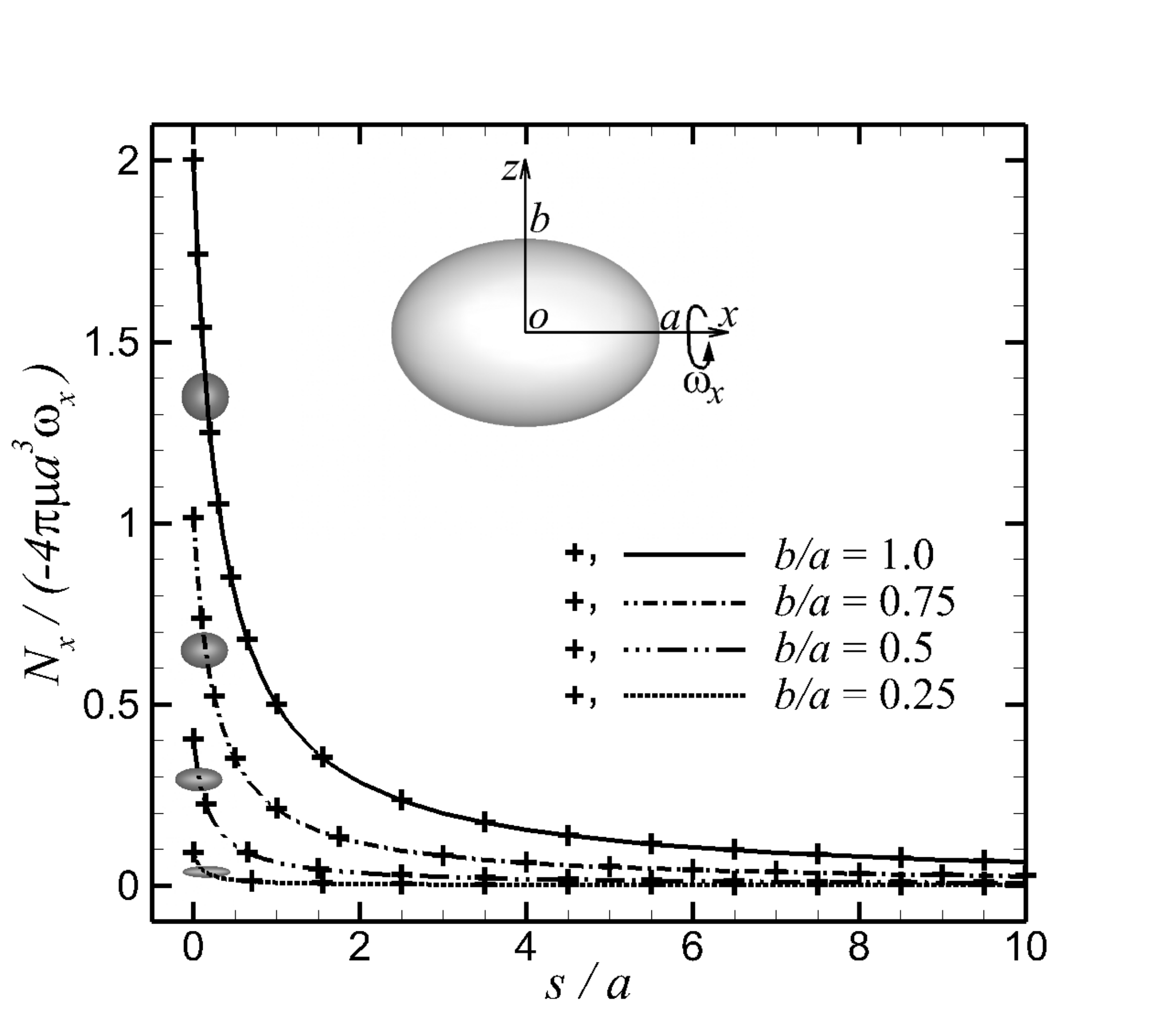} }
\subfloat[]{ \includegraphics[width=0.45\textwidth] {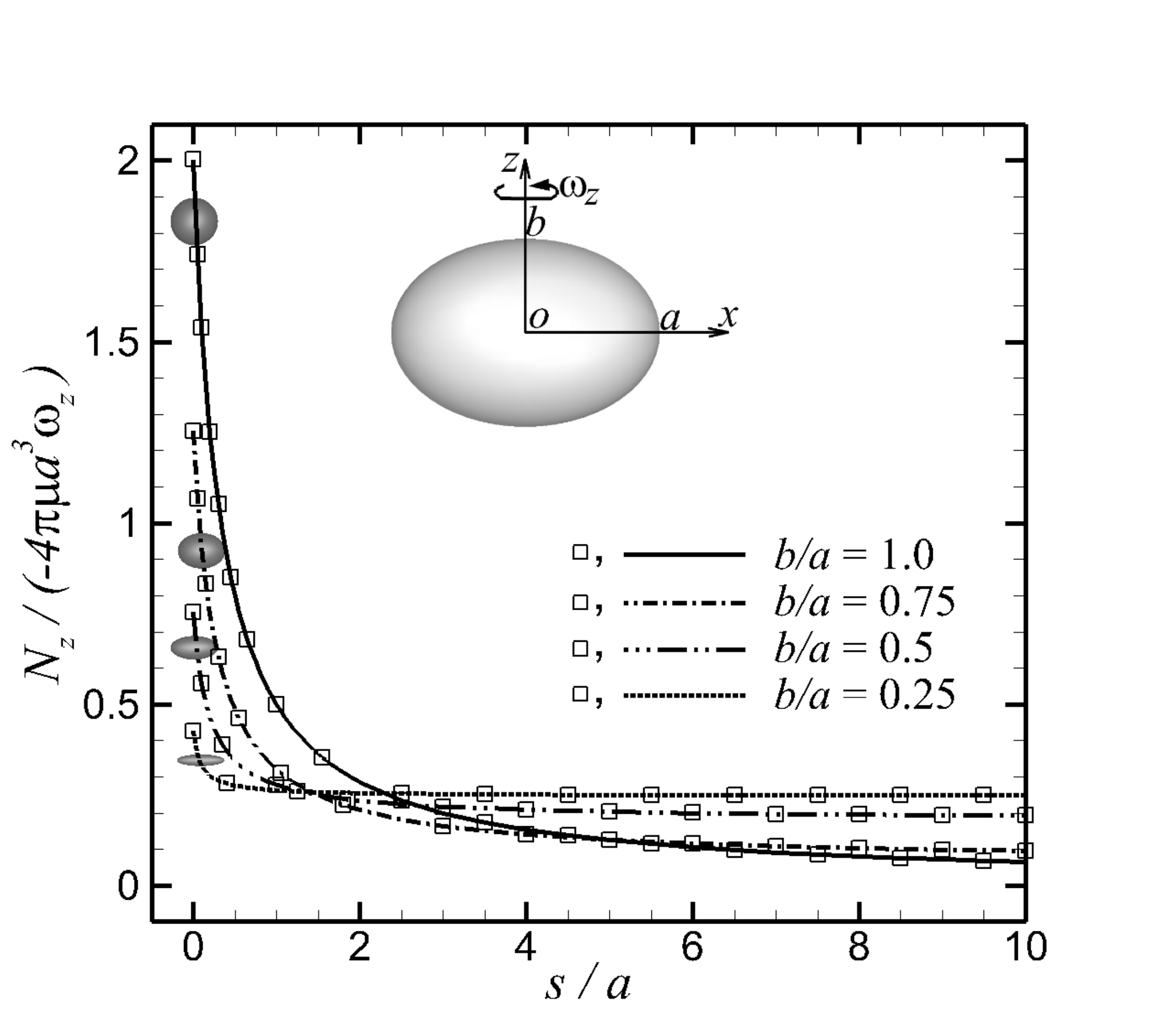} }
\caption{The torque coefficients in (a) $x$-direction and (b) $z$-direction of a rotating prolate spheroid with angular frequency $\omega_x$ and $\omega_z$ respectively as functions of $b/a$ in stationary Stokes flow; the torque coefficients in (c) $x$-direction and (d) $z$-direction of the rotating prolate spheroid as a function of the Navier slip length $s/a$ in stationary Stokes flow.  Results from BRIEF with the linear function for $\boldsymbol{w}$ (Item I in Table I) are shown as lines. For comparison, results shown as $(+)$ in (a) and (c) are obtained by BRIEF with the source potential $\boldsymbol{w}$ (Item IV in Table I), and results shown as $(\Box)$ in (b) and (d) are obtained by BRIEF with the stresslet $\boldsymbol{w}$ (Item V in Table I). Analytical results are shown as $(\bullet)$.\cite{Chwang1974} The BRIEF results with maximum relative error of less than 0.1\% are obtained using 2352 linear elements with 1178 nodes on the spheroid.}
\end{figure}}

We show that different auxiliary flow fields, $\boldsymbol{w}$, can be constructed using different known fundamental solutions of the Stokes equation as summarized in Eqs. (\ref{eq:wQ}), (\ref{eq:wS}) and Tables \ref{Tbl:M} and \ref{Tbl:D}. The absence of singular terms in the integrals means that even problems with surfaces that are very close together will not suffer any loss of numerical precision due to the adverse influence of the singularity of one surface upon a nearby surface. The regular nature of the BRIEF also provides a numerically robust way to evaluate field quantities near boundaries that is often more difficult to deal with than the singular behavior on the boundaries of the conventional boundary integral method.

By reformulating the boundary integral equation to remove the traditional singularities analytically, rather than developing integration algorithms to handle the integration over the element that contains the singular behavior, the same triangular mesh can be readily used to represent linear or quadratic elements. In practical implementations, the absence of any singular terms allows a significant reduction in the amount of computer code required with a corresponding reduction in the possibility of coding errors. And as we have seen, the use of quadratic elements offers a substantial gain in numerical precision. This flexibility therefore provides a convenient way to check the accuracy of a calculation by comparing results obtained from using linear and quadratic elements.

In all our examples, there is no practical difference in the numerical results obtained from using the different forms of the auxiliary function $\boldsymbol{w}$ given in Tables I and II. This is an expected result and indicates the analytical stability of BRIEF in removing the mathematical singularities in the conventional formulation of the boundary integral equations. Any convenient choice of $\boldsymbol{w}$ will result in a numerically robust scheme for solving the boundary integrals equation.

With all the above advantages, the boundary regularized integral equation formulation (BRIEF) of the Stokes problem presented here should always be used in preference to the traditional approach. Since the original physical fluid problem has no singularities, it is also intuitively satisfying that the corresponding numerical scheme can indeed be free of any singular behavior of a purely mathematical origin. Although we have only discussed application of the BRIEF to single-phase flow, an obvious development is to adopt this approach to multi-phase flow problems.

%==================================================
% -- Acknowledgments --
\begin{acknowledgments}
This work is supported in part by a Discovery Project Grant from the Australian Research Council to DYCC who is is a Visiting Scientist at the Institute of High Performance Computing and an Adjunct Professor at the National University of Singapore.
\end{acknowledgments}

\appendix
\section{Numerical implementation of BRIEF}
We now give the details of the numerical implementation of BRIEF of Eq. (\ref{eq:stkflnsbie}). Take the BRIEF with simply linear solution (Item I in Table I) for example, after introducing Eqs.~(\ref{eq:lnrw}) and (\ref{eq:lnrwtrF}) into Eq.~(\ref{eq:stkflnsbie}) by using the expression for $M_{jk}$ in Eq.~(\ref{eq:lnrMexprss}) that is listed at Item I of Table I, the boundary regularized integral formulation equation can be rearranged so that all velocities appear on the left-hand side and all tractions on the right-hand side
\begin{align}\label{eq:stknonsing2}
&\int_{S} (u_i-u_{i}^{0})T_{ijk}n_{k}\text{ d}S \nonumber\\
=  &\quad\text{ } \! \frac{1}{\mu}\! \int_{S} (f_i-f_{i}^{0}n^{0}_{l}n_{l}) {U}_{ij}\text{ d}S \nonumber \\
&+\frac{1}{\mu}\! \int_{S} \left[\frac{1}{2}(f_{k}^{0}n^{0}_{k})({\delta}_{il}+n_{i}^{0}n_{l}^{0})-f^{0}_{l}n_{i}^{0}\right]n_{l}{U}_{ij}\text{ d}S \nonumber \\
&+\frac{1}{\mu}\!\int_{S}\left[f_{i}^{0}n^{0}_{l}(x_{l}-x_{l}^{0})\right]{T}_{ijk}n_{k} \!\text{ d}S\nonumber\\
&+\frac{1}{\mu}\!\int_{S}\left[-\frac{1}{4}(f_{m}^{0}n^{0}_{m})({\delta}_{il}+n_{i}^{0}n_{l}^{0})(x_{l}-x_{l}^{0})
\right]{T}_{ijk}n_{k} \!\text{ d}S
\end{align}

\begin{figure}[t]
\centering
\includegraphics[width=0.87\textwidth]{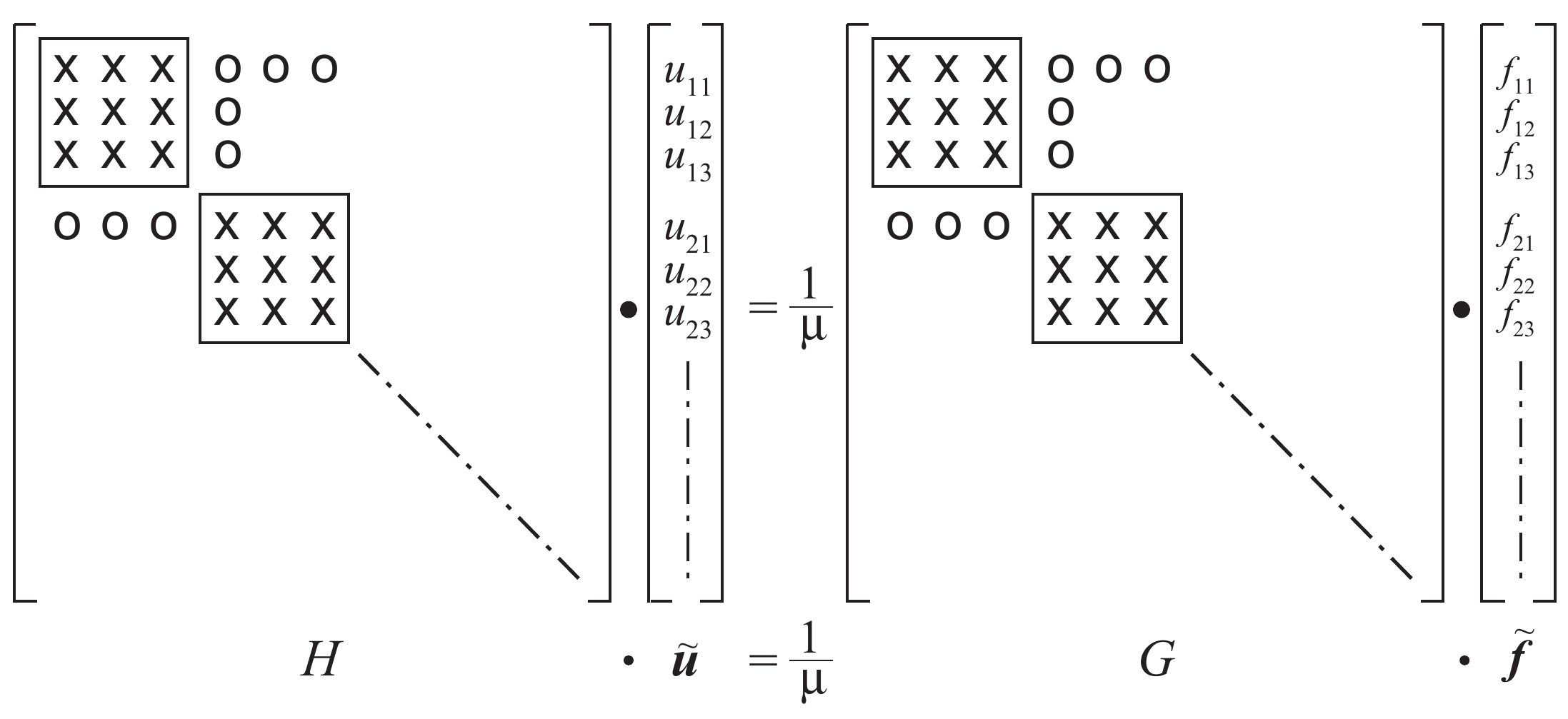}
\caption{Structure of the matrix equation corresponding to Eq.~(\ref{eq:stknonsing2}) for implementing the well-conditioned BIM. In the traditional BIM, the entries that appear as blocks of $3\times3$ in the matrices along the diagonal (indicated with an ``x'') will involve integrals over the singular point of the fundamental solutions.}\label{fig:stokeslnrmtrx}
\end{figure}

When the surface $S$ is partitioned into elements defined by $N$ nodes, the integrals in Eq.~(\ref{eq:stknonsing2}) over each element can be evaluated by quadrature and give rise to the matrix equation
\begin{align}\label{eq:stkmx}
\mathsfbi{H}\cdot \widetilde{\boldsymbol{u}} = \frac{1}{\mu} \mathsfbi{G} \cdot \widetilde{\boldsymbol{f}}
\end{align}
for the components of $\boldsymbol{u}$ and $\boldsymbol{f}$  at each node, see Fig~\ref{fig:stokeslnrmtrx}. Here, $\mathsfbi{H}$ is the $3N\times3N$ influence matrix corresponding to the velocity on the left-hand side of Eq.~(\ref{eq:stknonsing2}), and $\mathsfbi{G}$ is the $3N\times3N$ influence matrix corresponding to the traction on the right-hand side of Eq.~(\ref{eq:stknonsing2}). The vectors $\widetilde{\boldsymbol{u}}$ and $\widetilde{\boldsymbol{f}}$, both of size $3N$, contain every component of $\boldsymbol{u}$ and $\boldsymbol{f}$ at all the nodes. The coefficients corresponding to terms $u_{i}^{0}$ and $f_{i}^{0}$ appear in the ``x'' items of the $3\times3$ sub-matrices along the diagonals of the matrices $\mathsfbi{H}$ and $\mathsfbi{G}$, respectively, and the coefficients corresponding to terms $u_{i}$ and $f_{i}$ appear in the ``o'' items in the matrices $\mathsfbi{H}$ and $\mathsfbi{G}$, respectively.

%
%  Figure 14
%

In the traditional BIM formulation, the elements of the $3\times3$ sub-matrices along the diagonals of the matrices $\mathsfbi{H}$ and $\mathsfbi{G}$, indicated with an ``x'' within the square boxes in Fig.~\ref{fig:stokeslnrmtrx}, involved integrals over the singularities of the fundamental solutions. Numerous schemes have been developed to ensure the accurate numerical evaluation of such integrals~\cite{Gaul2003} and is one of the more arduous tasks in the numerical implementation of the BIM. But since the integrands in Eq.~(\ref{eq:stknonsing2}) are now free of singularities, all matrix entries can be obtained using, for example, the Gauss-Quadrature scheme to evaluate the integrals over each surface element.

To implement BRIEF with the linear solution for the auxiliary function $\boldsymbol{w}$, the number of lines of code needed to compute the matrix elements in squares that contain ``x'' along the diagonals of the matrices $\mathsfbi{H}$ and $\mathsfbi{G}$ is less than 40\% of that needed in the CBIM, where the local coordinate system transformation\cite{Gaul2003} is applied to deal with the singularities.

%==================================================
% -- References --
\bibliography{StkLetOn}% Produces the bibliography via BibTeX.

\end{document}